\RequirePackage{fix-cm}

\documentclass[twocolumn,10pt]{article}
\usepackage{etoolbox}
\usepackage{times}
\usepackage{authblk}
\usepackage[margin=2cm]{geometry}

\usepackage{amsmath}
\usepackage{amssymb}

\usepackage[utf8]{inputenc}

\usepackage{microtype}

\usepackage{balance}  

\usepackage{array}

\usepackage{tikz}
\usepackage{pgfplots}
\usepackage{soul} 

\usepackage{booktabs}

\usepackage{multirow}

\usepackage{xfrac}

\usepackage[ruled,vlined,linesnumbered,algo2e]{algorithm2e}

\DontPrintSemicolon
\SetAlFnt{\small}

\SetCommentSty{mycommfont}
\SetAlgorithmName{\hspace*{-0.6em}Algorithm}

\SetKwInOut{KwData}{input}
\SetKwInOut{KwResult}{output}

\SetKw{And}{and}
\SetKw{Or}{or}
\SetKw{Not}{not}
\SetKw{Var}{var}
\SetKw{New}{new}

\SetKwData{Consumer}{consumer}
\SetKwData{IdxR}{idxR}
\SetKwData{IdxS}{idxS}
\SetKwData{ItR}{itR}
\SetKwData{ItS}{itS}
\SetKwData{ActiveR}{activeR}
\SetKwData{Buffer}{buffer}
\SetKwData{Comp}{comp}
\SetKwFunction{Insert}{insert}
\SetKwFunction{Erase}{erase}
\SetKwFunction{Remove}{remove}
\SetKwFunction{GetEndpoint}{getEndpoint}
\SetKwFunction{MoveToNextEndpoint}{moveToNextEndpoint}
\SetKwFunction{IsFinished}{isFinished}
\SetKwFunction{Clear}{clear}
\SetKwFunction{IsFull}{isFull}
\SetKwProg{Function}{function}{}{end}

\colorlet{DanAddColor}{green!40}
\colorlet{DanDelColor}{red!27}

\newcommand*{\Tuple}[1]{\ensuremath{\langle #1\rangle}}

\newcommand*{\Set}[1]{\ensuremath{\{#1\}}}
\newcommand*{\SetBuilder}[2]{\Set{#1\mid#2}}

\newcommand*{\Rel}[1]{\ensuremath{\boldsymbol{#1}}}

\newcommand*{\JoinName}[1]{{\textsc{\lowercase{#1}}}}
\newcommand*{\RelationName}[1]{\JoinName{#1}}

\newcommand*{\Size}[1]{|#1|}

\newcommand*{\BigO}{\ensuremath{O}}

\newcommand*{\RefPage}[1]{\ref{#1} (page~\pageref{#1})}
\newcommand*{\RefOnPage}[1]{\ref{#1} on page~\pageref{#1}}

\renewcommand*{\leq}{\leqslant}
\renewcommand*{\le} {\leqslant}
\renewcommand*{\geq}{\geqslant}
\renewcommand*{\ge} {\geqslant}

\let\OldLAnd\land
\renewcommand*{\land}{\mathrel{\operatorname{\OldLAnd}}}

\makeatletter
\newcommand*\bigcdot{\mathpalette\bigcdot@{.5}}
\newcommand*\bigcdot@[2]{\mathbin{\vcenter{\hbox{\scalebox{#2}{$\m@th#1\bullet$}}}}}
\makeatother

\newcommand*{\JoinByS}{\mathrel{\operatorname{
    \raisebox{-0.6pt}{\scalebox{1.3}{$\Join\hspace{-0.595em}\bigcdot\hspace{0.04em}$}}
}}}

\newcommand*{\JoinBySsm}{\mathrel{\operatorname{
    \raisebox{-0.6pt}{\scalebox{1.3}{$\Join\hspace{-0.535em}\bigcdot\hspace{0.04em}$}}
}}}

\newcommand*{\JoinByTs}{\JoinByS_{\mathrm{start}}}
\newcommand*{\JoinByTe}{\JoinByS_{\mathrm{end}}}

\newcommand*{\JoinByTssm}{\JoinBySsm_{\mathrm{start}}}
\newcommand*{\JoinByTesm}{\JoinBySsm_{\mathrm{end}}}

\newcommand*{\Map}[2]{\chi_{#1:#2}}

\newcommand*{\Select}{\sigma}

\newcommand*{\AfterTableCaptionSpace}{\vspace{-1.5mm}}

\newcommand{\url}[1]{\texttt{#1}}

\graphicspath{{images/}}

\usepackage{tikz}
\usetikzlibrary{arrows.meta}
\usetikzlibrary{calc}

\colorlet{MyGreen}{green!50!black}

\pgfdeclarelayer{shading}
\pgfdeclarelayer{grid}
\pgfdeclarelayer{overlay}
\pgfdeclarelayer{axes}
\pgfdeclarelayer{tuples}
\pgfsetlayers{shading,grid,overlay,tuples,axes,main}

\tikzstyle{relations}       = [yscale = 0.4]
\tikzstyle{axis}            = []
\tikzstyle{axis label}      = [font = \scriptsize, inner sep = 0, outer sep = 3pt]
\tikzstyle{x axis label}    = [axis label, anchor = 110]
\tikzstyle{y axis label}    = [axis label, anchor = -20]
\tikzstyle{axis tick}       = [thin]
\tikzstyle{axis tick label} = [font = {\normalfont\scriptsize}, inner sep = 0, outer sep = 4pt]
\tikzstyle{grid}            = [thin, lightgray]
\tikzstyle{tuple}           = [line width = 1mm]
\tikzstyle{old tuple}       = [tuple, gray, dash pattern = on 1mm off 1mm]
\tikzstyle{tuple label}     = [font = \small, at start, anchor = south west,
    inner ysep = 1.5pt, inner xsep = 1pt]
\tikzstyle{overlay line}    = [orange, line width = 2mm]

\newcommand*{\DrawXAxis}[3] 
{
    \pgfmathsetmacro{\MaxX}{#1}
    \pgfmathsetmacro{\MaxY}{#2}

    \begin{pgfonlayer}{grid}
        \ifnum#3>0
        \draw [grid] (0, #3) -- (\MaxX, #3);
        \fi
        \foreach \x in {0, ..., \MaxX}
        \draw [grid] (\x, 0) -- +(0, \MaxY);
    \end{pgfonlayer}

    \begin{pgfonlayer}{axes}
        \draw[axis, ->] (0, 0) -- (\MaxX, 0) node [x axis label] {$t$};
        \pgfmathparse{\MaxX - 1}
        \foreach \x in {0, ..., \pgfmathresult}
        {
        \draw[axis tick] (\x, -0.2) -- (\x, 0.2);
        \node[axis tick label, anchor = north] at (\x, 0) {\x};
        }
    \end{pgfonlayer}
}

\newcommand*{\DrawAxes}[2] 
{
\pgfmathsetmacro{\MaxX}{#1}
\pgfmathsetmacro{\MaxY}{#2}

\DrawXAxis{\MaxX}{\MaxY}{0}

\begin{pgfonlayer}{grid}
    \foreach \y in {1, ..., \MaxY}
    \draw [grid] (0, \y) -- +(\MaxX, 0);
\end{pgfonlayer}

\begin{pgfonlayer}{axes}
    \draw[axis, ->] (0, 0) -- (0, \MaxY) node [y axis label] {$A$} ;
    \pgfmathparse{\MaxY - 1}
    \foreach \y in {0, ..., \pgfmathresult}
    {
    \draw[axis tick] (-0.1, \y) -- (0.1, \y);
    \node[axis tick label, anchor = east] at (0, \y) {\y};
    }
\end{pgfonlayer}
}

\newcommand*{\DrawTuple}[5][tuple] 
{
    \begin{pgfonlayer}{tuples}
        \draw[#1] (#2, #4) -- (#3, #4) node [tuple label] {#5};
    \end{pgfonlayer}
}

\newcommand*{\DrawRTuple}{}
\newcommand*{\ResetDrawRTuple}{
    \renewcommand*{\DrawRTuple}[4]{\DrawTuple{##1}{##2}{##3}{##4}}
}
\ResetDrawRTuple

\newcommand*{\DrawISEQLExampleRelations}
{
    \DrawXAxis{8}{7}{3}

    \DrawRTuple{0}{1}{5}{$r_1$}
    \DrawRTuple{1}{3}{6}{$r_2$}
    \DrawRTuple{2}{5}{4}{$r_3$}

    \DrawTuple{1}{3}{2}{$s_1$}
    \DrawTuple{3}{4}{1}{$s_2$}
}

\newsavebox{\IgnoreBox}
\tikzset
{
    ignore/.style = 
    {
        draw = none,
        fill = none,
        overlay,
        execute at begin node = {\begin{lrbox}{\IgnoreBox}},
        execute at end node = {\end{lrbox}}
    },
    table interval algebra relation/.style =
    {
        xscale = 0.5,
        yscale = 0.2,
        interval/.style = {line width = 1pt, > = {Circle[length = 3.4pt]},
            shorten > = -0pt},
        baseline = {($(current bounding box.north) - (0,7.5pt)$)},
        interval label/.style = {font = {\small}},
        vertical separator/.style = {very thin, overlay},
    },
    inline interval algebra relation/.style =
    {
        xscale = 0.1,
        yscale = 0.1,
        baseline = -1pt,
        interval/.style = {line width = 0.5pt, > = {Circle[length = 1.4pt]},
            shorten > = 0pt},
        interval label/.style = {ignore},
        vertical separator/.style = {ignore},
    },
}

\newcommand*{\Doodle}[2][inline] 
{%
    \begin{tikzpicture}[#1 interval algebra relation]
        #2
    \end{tikzpicture}%
}

\newcommand*{\Interval}[5][] 
{
    \draw[interval, #1] (#2, #4) -- (#3, #4);
    \node[interval label] at (-0.5,#4) {#5};
}

\newcommand*{\RSIntervals}[6] 
{
    \Interval[#1]{#2}{#3}{1}{$r$}
    \Interval[#4]{#5}{#6}{0}{$s$}
}

\newcommand*{\Reverse}[1]
{
    \begin{scope}[yscale = -1, yshift = -1cm]
      #1
    \end{scope}
}

\newcommand*{\LeftOverlap}
{
    \RSIntervals
        {<->}{0}   {2}
        {<->}   {1}   {3}
}

\newcommand*{\During}
{
    \RSIntervals
        {<->}   {1}{2}
        {<->}{0}      {3}
}

\newcommand*{\StartPreceding}
{
    \RSIntervals
        {<- }{0}      {3}
        {<- }   {1}   {3}
}

\newcommand*{\EndFollowing}
{
    \RSIntervals
        { ->}{0}      {3}
        { ->}{0}   {2}
}

\newcommand*{\Before}
{
    \RSIntervals
        { ->}{0}{1}
        {<- }      {2}{3}
}

\newcommand*{\Meets}
{
    \RSIntervals
        { ->}{0}{1.495}
        {<- }        {1.505}{3}
    \draw [vertical separator] (1.5, -0.5) -- (1.5, 1.5);
}

\newcommand*{\Starts}
{
    \RSIntervals
        {<->}{0}{1.5}
        {<->}{0}     {3}
}

\newcommand*{\Finishes}
{
    \RSIntervals
        {<->}   {1.5}{3}
        {<->}{0}     {3}
}

\newcommand*{\Equals}
{
    \RSIntervals
        {<->}{0}     {3}
        {<->}{0}     {3}
}

\newcommand*{\Overlaps}{\LeftOverlap}

\pgfplotsset{compat = 1.12}

\usepackage{pgfplotstable}
\usepgfplotslibrary{groupplots}

\pgfplotscreateplotcyclelist{mylist}
{
    {red,     mark = *,         mark size = 1.92},
    {blue,    mark = square*,   mark size = 1.66},
    {MyGreen, mark = triangle*, mark size = 2.13},
    {brown,   mark = diamond*,  mark size = 2.48},
    {red,     mark = star,      mark size = 2.10},
    {blue,    mark = square,    mark size = 1.66},
    {MyGreen, mark = triangle,  mark size = 2.13},
    {brown,   mark = diamond,   mark size = 2.48},
}

\pgfplotsset
{
    global plot style/.style = 
    {
        scale only axis,
        width = 74mm,    
        height = 2.5cm,   
        ymajorgrids,
        legend style = 
        {
            at = {(0.5, 1.05)}, 
            anchor = 270,
            /tikz/every even column/.append style = {column sep = 1mm},
        },
        max space between ticks = 20pt,
        legend columns = -1,     
        cycle list name=mylist,
        tick label style  = {font = \scriptsize, inner sep = 0pt, outer sep = 3pt},
        x tick label style  = {},
        label style       = {font = \footnotesize, inner sep = 0pt, outer sep = 7pt},
        legend style      = {font = \scriptsize, draw=none, fill=none},
        title style       = {font = \scriptsize, inner sep = 0, yshift = -0.4mm},    
    },
    default plot style/.style = {
         /pgf/number format/1000 sep={\,},
         ytickten = {-1, 0, ..., 5},
         grid style = dotted,
         minor grid style = dotted,
         ymajorgrids,
         legend pos = north west,
    },
    log y ticks with fixed point/.style = {
         yticklabel=
         {
               \pgfkeys{/pgf/fpu=true}
               \pgfmathparse{exp(\tick)}%
               \pgfmathprintnumber[fixed relative, precision=3]{\pgfmathresult}
               \pgfkeys{/pgf/fpu=false}
          }
    },
    minor y grid/.style = {
          grid style = solid,
          minor grid style = dotted,
          ymajorgrids, yminorgrids,
    },
    three plots/.style = {
        global plot style,
        group/rows = 1,
        group/columns = 3,
        group/ylabels at = edge left,
        group/xlabels at = edge bottom,
        group/horizontal sep = 2mm,
        group/vertical sep = 12mm,
        width = 50mm, 
        height = 25mm,
        multiple plots middle/.style = {
            yticklabels = \empty,
        },
        multiple plots right/.style = {
            yticklabel pos = right,
        },
    },
    four plots/.style = {
        three plots,
        group/columns = 4,
        width = 40mm, 
    },
    real world bar plot/.style = {
        symbolic x coords = {flight,inc,web,basf,feed,wi,fi},
        xtick = data,
        ybar, /pgf/bar width = 5pt, ybar = 0pt,
        xtick style={draw=none},
       	xlabel = {Real-world dataset},
    },
    real-world join bar plot/.style = {
                global plot style,
                cycle list shift=-2, 
                xbar, area legend, reverse legend, bar width = 6pt, height=10mm,
                symbolic y coords={pre, #1, after}, ymin={pre}, ymax={after},
                xtick={0}, xmin = 0, enlarge x limits={upper, value=0.2},
                ytick=data, y tick label style={rotate=90,anchor=south},
                legend style={at={(0.5,1.1)}, anchor=south, legend columns=1},
                xlabel = {Joining time, s (linear scale)},
                nodes near coords, nodes near coords align={horizontal},every node near coord/.append style={font=\scriptsize},
                /pgf/number format/fixed,
                scaled y ticks = false,
                scaled x ticks = false
    },
    real-world join bar plot small/.style = {
                real-world join bar plot = {#1},
                height = 0.8cm,
                width=7cm,
                cycle list shift=-1,
                ymajorgrids = false,
    },
    list performance plot/.style = {
         xmin=1e2, xmax=1e8,     
         default plot style,     
         minor y grid,           
         log y ticks with fixed point,
         xlabel = {Number of tuples},   
         height = 3.8cm,
         width = 8cm,
         legend style = 
         {
             draw=none, 
             fill=none, 
             font=\scriptsize,       
             at = {(0.5, 1.05)}, 
             anchor = 270,
             /tikz/every even column/.append style = {column sep = 1mm},
         },
         legend columns = -1,     
         cycle list name=mylist,
    },  
    list performance plot small/.style = {
         list performance plot,
         legend style={font = \scriptsize, draw=none, fill=none},
         height = 3.8cm,
         width = 5.35cm,
         xtickten = {2,3,...,8},
    },
    log y ticks with fixed point base 2/.style = {
        yticklabel = {
            \pgfkeys{/pgf/fpu=true}
            \pgfmathparse{2^\tick)}%
            \pgfmathprintnumber[fixed relative, precision=3]{\pgfmathresult}
            \pgfkeys{/pgf/fpu=false}
        }
    },
    small synthetic join plot base/.style =
    {
            ymin = 1e0, ymax = 8e4,
         legend style={draw=none, fill=none},                       
            default plot style,
            ylabel = {Joining time, s},
            height = 1.8cm,
    },
    small synthetic join plot/.style =
    {
            small synthetic join plot base,
            symbolic x coords={u1,u2,u3,u4,u5,u6,u7,u8,u9,{$e_1$},{$e_2$},{$e_3$},{$e_4$},{$e_5$},{$e_6$},{$e_7$},{$e_8$},{$e_9$}, {$z_0$}, {$z_1$}, {$z_2$}, {$z_3$}, {$z_4$}, {$z_5$}, {$z_6$}, {$z_7$}},
            xmin = {$e_2$}, xmax = {$e_8$},
            xlabel = {Workload},
    },
}

\newcommand*{\ReadTable}[1]{\pgfplotstableread{data/#1}\data}
\newcommand*{\AddPlot}[2][]{\addplot+[#1] table [#2] from \data;}

\newcommand*{\AddPlotCoord}[1]{
\pgfmathsetmacro\tmpval{\pgfplotsretval}
\addplot coordinates {(\tmpval,#1)};
}

\pgfplotstableread{data/hash-sort-merge-interval-join-data-p-benchmark-lists.txt}\dataPLists
\pgfplotstableread{data/hash-sort-merge-interval-join-data-zipf.txt}\dataZipf
\pgfplotstableread{data/hash-sort-merge-interval-join-data-rw.txt}\dataRW
\pgfplotstableread{data/hash-sort-merge-interval-join-data-fs.txt}\dataFS

\newbool{CompileTechReport}
\newcommand{\ifnottechreport}[1]{\ifbool{CompileTechReport}{}{#1}}
\newcommand{\iftechreport}[1]{\ifbool{CompileTechReport}{#1}{}}

\booltrue{CompileTechReport}

\newtheorem{example}{Example}
\newtheorem{definition}{Definition}
\renewcommand*{\AfterTableCaptionSpace}{}

\ifnottechreport{
\title
{Cache-Efficient Sweeping-Based Interval Joins
for Extended Allen Relation Predicates}
}

\iftechreport{
\title
{Cache-Efficient Sweeping-Based Interval Joins
for Extended Allen Relation Predicates (Extended Version)}
}

\author[1]{Danila Piatov}
\author[2]{Sven Helmer}
\author[1]{Anton Dign{\"o}s}
\author[1]{Fabio Persia}
\affil[1]{Free University of Bozen-Bolzano, Bozen-Bolzano, Italy}
\affil[ ]{\url{firstname.lastname@unibz.it}}
\affil[2]{University of Zurich, Zurich, Switzerland}
\affil[ ]{\url{helmer@ifi.uzh.ch}}
\date{}                     
\usepackage{calc}
\usepackage{xcolor,colortbl}
\newcommand{\REV}[1]{{#1}}
\definecolor{Gray}{gray}{0.98}

\begin{document}

\maketitle

\begin{abstract}
We develop a family of efficient plane-sweeping interval join algorithms that
can evaluate a wide range of interval predicates such as Allen's relationships
and parameterized relationships. Our technique is based on a framework,
components of which can be flexibly combined in different manners to support
the required interval relation. In temporal databases, our algorithms can
exploit a well-known and flexible access method, the Timeline Index, thus
expanding the set of operations it supports even further. Additionally,
employing a compact data structure, the gapless hash map, we
utilize the CPU cache efficiently. In an experimental evaluation, we show
that our approach is several times faster and scales better than
state-of-the-art techniques, while being much better suited for real-time
event processing.
\end{abstract}


\section{Introduction}
\label{sec:introduction}

Temporal data is found in many financial, business, and scientific
applications running on top of database management systems (DBMSs), i.e.,
supporting these applications through efficient temporal operator
implementations is crucial. For example, Kaufmann states that there
are several temporal queries in the hundred most expensive queries executed on
SAP ERP~\cite{Kauf14}, many of which have to be implemented in the application
layer, as the underlying infrastructure does not directly support the
processing of temporal data.  According to~\cite{Kauf14}, customers of SAP
desperately need (advanced) temporal operators for efficiently running queries
pertaining to legal, compliance, and auditing processes.

Although the introduction of temporal operators into the SQL standard has
started with SQL:2011~\cite{DBLP:journals/sigmod/KulkarniM12}, the provided
implementation is far from complete or lacking in performance. There
is a renewed interest in temporal data processing, and researchers and
developers are busy filling the gaps. One example are join operators involving
temporal predicates: there are several recent publications on overlap interval
joins~\cite{bouros_forward_2017,dignos_overlap_2014,piatov_interval_2016}. However,
this is not the only possible join predicate for matching (temporal)
intervals. Allen defined a set of binary relations between intervals
originally designed for reasoning about intervals and interval-based temporal
descriptions of events~\cite{allen_maintaining_1983}. These relations have
been extended for event detection by parameterizing
them~\cite{helmer_iseql_2016}. Strictly speaking, all these relationships
could be formulated in regular SQL \texttt{WHERE} clauses (see also the
right-hand column of Table~\ref{table:allen-relations} for a formal definition of Allen's relations
and extensions). The evaluation of these predicates using the implementation
present in contemporary relational database management systems (RDBMSs) would
be very inefficient, though, as a lot of inequality predicates are involved
\cite{khayyat_fast_2017}.  We also note that the predicates in
the aforementioned overlap interval joins
(\cite{bouros_forward_2017,dignos_overlap_2014,piatov_interval_2016}) only
check for any form of overlap between intervals. Basically, they do not
distinguish between many of the relationships defined by Allen and do not
cover the \RelationName{Before} and \RelationName{Meets} relations at all.
Additionally, many of the approaches so far lack parameterized versions, in
which further range-based constraints can be formulated directly in the join
predicate.

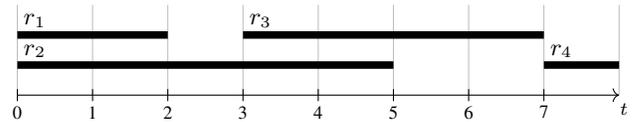
\begin{figure}[t]
\centering
\begin{tikzpicture}[relations]
    \DrawXAxis{8}{3}{0}
    \DrawTuple{0}{2}{2}{$r_1$}
    \DrawTuple{0}{5}{1}{$r_2$}
    \DrawTuple{3}{7}{2}{$r_3$}
    \DrawTuple{7}{8}{1}{$r_4$}
\end{tikzpicture}
\caption{Example temporal relation \Rel r}
\label{fig:employees}
\vspace*{-.19cm}
\end{figure}

In Figure~\ref{fig:employees} we see an example relation showing which employees ($r_1$, $r_2$, $r_3$, and $r_4$) were working on a certain project during which month.
The tuple validity intervals (visualized as line segments) are as follows:
tuple $r_1$ is valid from time 0 to time 2 (exclusive) with a starting timestamp $T_s = 0$ and an ending timestamp $T_e = 2$, tuple $r_2$ is valid on interval $[0, 5)$ with $T_s = 0$ and $T_e = 5$, and so on.
With a simple overlap interval join we can merely detect that $r_1$ and $r_2$ worked together on the project for some time (as did $r_2$ and $r_3$).
However, we may also be interested in who started working at the same time ($r_1$ and $r_2$), who started working after someone had already left ($r_3$ coming in after $r_1$ had left and $r_4$ starting after everyone else had left), or even who took over from someone else, i.e., the ending of one interval coincides with the beginning of another one ($r_3$ and $r_4$).
For even more sophisticated queries, we may want to add thresholds: who worked together with someone else and then left the project within two months of the other person ($r_3$ and $r_2$).

\REV{
Allen's relationships are not only used in temporal databases, but also in
event detection systems for temporal pattern
matching~\cite{helmer_iseql_2016,KGMS19,KGS18}. In this context, it is also
important to be able to specify concrete timeframes within which certain
patterns are encountered, introducing the need for parameterized versions of
Allen's relationships. For instance, K{\"o}rber et al. use their TPStream
framework for analyzing real-time traffic data~\cite{KGMS19,KGS18}, while we
previously employed a language called ISEQL to specify events in a video
surveillance context~\cite{helmer_iseql_2016}.}
Event detection motivated
us to develop an approach that is also applicable for event stream processing
environments, meaning that our join operators are non-blocking and produce
output tuples as early as logically possible, without necessarily waiting for the intervals to finish.
Moreover, we demonstrate how these joins can be
processed efficiently in temporal databases using a sweeping-based framework that
is supported by cache-efficient data structures and by a Timeline Index --- a flexible and general
temporal index --- supporting a wide range of temporal operators, used in a
prototype implementation of SAP HANA \cite{kaufmann_timeline_2013}.
We therefore extend the set of operations Timeline Index supports,
increasing its usefulness even further.


In particular, we make the following contributions:

\begin{itemize}

\item We develop a family of plane-sweeping interval join algorithms
  that can evaluate a wide range of interval relationship predicates going
  even beyond Allen's relations.

\item At the core of this framework sits one base algorithm, called
  interval-timestamp join, that can be
  parameterized using a set of iterators trav\-ersing a Timeline Index. This
  offers an elegant way of configuring and adapting the base algorithm
  for processing different interval join predicates, \REV{improving code maintainability}.

\item Additionally, our algorithm utilizes the CPU cache efficiently, relying
  on a compact hash map data structure for managing data during the processing
  of the join operation. Together with the index, in many cases we can achieve
  linear run time. 
  
\item In an experimental evaluation, we show that our approach is faster than
  state-of-the-art methods: an order of magnitude faster than a direct
  competitor and several orders of magnitude faster than an inequality join.

\end{itemize}


\section{Related Work}

There is a renewed interest in employing Allen's interval relations in
different areas, e.g. for describing complex temporal events in event
detection frameworks~\cite{helmer_iseql_2016,KGMS19,KGS18} as well as for querying
temporal relationships in knowledge graphs via SPARQL~\cite{CPS19}. One reason
is that it is more natural for humans to work with chunks of information, such as
labeled intervals, rather than individual values~\cite{HoPe14}.

\subsection{Allen's Interval Relations Joins}
\label{sec:leung-muntz}

Leung and Muntz worked on efficiently implementing joins with predicates based
on Allen's relations in the 1990s
\cite{leung_query_1989,leung_query_1990} and it turns out that their
solution is still competitive today. In fact, they also apply a
plane-sweeping strategy, but impose a total order on the tuples of a
relation. Theoretically, there are four different orders tuples can be sorted
in for this algorithm:
$T_s$ ascending, $T_s$ descending, $T_e$ ascending, and $T_e$
descending. When joining two relations, they can be sorted in different orders
independently of each other.

The actual algorithm is similar to a sort-merge join. A tuple is read
from one of the relations (outer or inner) and placed into the corresponding
set of active tuples for that relation. Each tuple in the set of the other
relation is checked whether it matches the tuple that was just read. When a
matching pair is found, it is transferred to the result set. While searching
for matching tuples, the algorithm also performs a garbage collection,
removing tuples that will no longer be able to find a matching
partner. (Not all join predicates and sort orders allow for a garbage
  collection, though.) A heuristic, based on tuple distribution and garbage collection
statistics, decides from which relation to read the next tuple.
In a follow-up paper, further strategies for parallelization and temporal
query processing are discussed~\cite{leung_temporal_1992}.

In contrast to our approach, in which we handle tuple starting and ending
events separately (an idea also covered more generally 
in~\cite{Freksa92,MaHa06,Wuch09}), the algorithm of Leung and Muntz requires
streams of whole tuples. A tuple is not complete until its ending endpoint
$T_e$ is encountered. This has a major impact for applications such as
real-time event detection. Waiting for a tuple to finish can delay the whole
joining process, as tuples following it in the sort order cannot be
reported yet.

Chekol et al. claim to cover the complete set of Allen's relations in their
join algorithm for intervals in the context of SPARQL, but the description for
some relations is missing~\cite{CPS19}. It seems they are using our algorithm
from~\cite{piatov_interval_2016} as a basis. They are not able to handle
parameterized versions and have to create different indexes for different
relations, though.

There is also research on integrating Allen's predicate interval
joins in a MapReduce
framework~\cite{chawda_processing_2014,pilourdault_distributed_2016}.
However, these approaches focus on the effective distribution of the data over
MapReduce workers rather than on effective joins.

\subsection{Overlap Interval Joins}

One of the earliest publications to look at performance issues of temporal
joins is by Segev and Gunadhi~\cite{segev_event-join_1989,gunadhi_query_1991}, who compare different
sort-merge and nested-loop implementations of their event join. They refined existing
algorithms by applying an auxiliary access method called an append-only tree,
assuming that temporal data is only appended to existing relations and never
updated or deleted.

Some of the work on spatial joins can also be applied to interval joins.
Arge et~al.~\cite{arge_scalable_1998} used a sweeping-based interval join algorithm
as a building block for a two-dimensional spatial rectangle join, but did not
investigate it as a standalone interval join. It was picked up again by Gao et
al.~\cite{gao_join_2005}, who
give a taxonomy of temporal join operators and provide a survey and an empirical
study of a considerable number of non-index-based temporal join algorithms,
such as variants of nested-loop, sort-merge, and partitioning-based
methods.

The fastest partitioning join, the overlap interval
partitioning (OIP) join, was developed by Dign\"os et al.
\cite{dignos_overlap_2014}. The
(temporal) domain is divided into equally-sized granules and adjacent
granules can be combined to form containers of different sizes. Intervals are
assigned to the smallest container that covers them and the join algorithm
then matches intervals in overlapping containers.


The Timeline Index, introduced by Kaufmann
et~al.~\cite{kaufmann_timeline_2013}, and the supported
temporal operators have also received
renewed attention recently. Kaufmann et~al. showed that
a single index per temporal relation supports such operations as time travel,
temporal joins, overlap interval joins and temporal aggregation on constant
intervals (temporal grouping by instant). The work was done in the context of
developing a prototype for a commercial temporal database and later
extended to support bitemporal data~\cite{kaufmann_bi-temporal_2015}.

In earlier work~\cite{piatov_interval_2016}, we defined a simplified, but
functionally equivalent version of the Timeline Index, called Endpoint Index
(from now on we will use both terms interchangeably). We introduced a
cache-optimized algorithm for the overlap interval join, based on the Endpoint
Index, and showed that the Timeline Index is not only a universal index
that supports many operations, but that it can also outperform the
state-of-the-art specialized indexes (including~\cite{dignos_overlap_2014}).
The technique of sweeping-based algorithms has recently been applied to
temporal aggregation as well~\cite{BoMa18,piatov_sweeping-based_2017}, extending the
set of operations supported by the Timeline Index even further.

\REV{
The basic idea of Bouros and Mamoulis is to do a \emph{forward
  scan} on the input collection to determine the join partners, hence their
algorithm is called forward scan (FS) join~\cite{bouros_forward_2017}.
In contrast, our approach does a
\emph{backward scan} by traversing already encountered intervals, which have
to be stored in a hash table. In FS, both input relations, $\Rel r$ and $\Rel s$,
are sorted on the starting endpoint of each interval and then the algorithm
sweeps through the endpoints of $\Rel r$ and $\Rel s$ in order. Every time in
encounters an endpoint of an interval, it scans the other relation and joins
the interval with all matching intervals. Bouros and Mamoulis introduce a
couple of optimizations to improve the performance. First, consecutively swept
intervals are \emph{grouped} and processed in batch (this is called
gFS). Second, the (temporal) domain is split into tiles and intervals starting
in such a tile are stored in a \emph{bucket} associated with this tile. While
scanning for join partners for a tuple $r$, all intervals in buckets
corresponding to tiles that are completely covered by the interval of $r$ can
be joined without further comparisons. Combined with the previous technique,
this results in a variant called bgFS. Doing a forward scan or a backward scan
has certain implications. Introducing their optimizations to FS, Bouros and
Mamoulis showed that forward scanning is usually more efficient than backward
scanning (particularly when it comes to parallelizing the algorithm). However,
there is also a downside: forward scanning needs to have access to the
complete relations to work, while backward scanning considers only already
encountered endpoints, i.e., backward scanning can be utilized in a streaming
context (for forward scanning this is not possible).
}

\subsection{Generic Inequality Joins} 

\REV{As we will see in Table~\ref{table:allen-relations}, most of the interval
joins can be broken down into inequality joins, which becomes very inefficient
as soon as more than one inequality predicate is involved: Khayyat et
al.~\cite{khayyat_fast_2017} point out that these joins are handled via
naive nested-loop joins in contemporary RDBMSs. They develop a more efficient
inequality join (IEJoin), which first sorts the relations according to the
join attributes. For the sake of simplicity, we just consider two inequality
predicates here, i.e., for every relation \Rel r, we have two versions, \Rel
{r^1} and \Rel {r^2} sorted by the two join attributes, which helps us to
find the values satisfying an inequality predicates more efficiently. (The
connections between tuples from \Rel {r^1} and \Rel {r^2} are made using a
permutation array.) Some additional data structures, offset
arrays and bit arrays, help the algorithm to take shortcuts, but
essentially the basic join algorithm still consists of two nested loops,
leading to a quadratic run-time complexity (albeit with a performance that is
an order of magnitude better than a naive nested-loop join).}

%
%
%

\section{Background}


\paragraph*{Interval Data:\,} 

We define a \emph{temporal tuple} as a relational tuple containing two
additional attributes, $T_s$ and $T_e$, denoting the start and end of the
half-open tuple validity interval $T = [T_s, T_e)$.
We will use a period (.) to denote an attribute of a tuple, e.g. $r.T_s$ or
$s.T$. The length of the tuple validity interval, $\Size r$, is therefore
$r.T_e - r.T_s$. We use the terms \emph{interval} and \emph{tuple}
interchangeably. With $r$ and $s$ we denote the
left-hand-side and the right-hand-side tuples in a join, respectively.
\REV{We use integers for the timestamps to simplify the explanations. Our
  approach would work with any discrete time domain: we require a total order
  on the timestamps and a fixed granularity, i.e., given a timestamp we have
  to be able to unambiguously determine the following one.}


\paragraph*{Interval Relations:\,} 

As intervals accommodate the human perception of time-based patterns much
better than individual values~\cite{HoPe14}, intervals and their relationships are a
well-known and widespread approach to handle temporal data~\cite{Jensen99}. Here we
look at two different ways to define binary relationships between intervals:
Allen's relations~\cite{allen_maintaining_1983} and the Interval-based Surveillance
Event Query Language (ISEQL) \cite{bettini_2017,helmer_iseql_2016}.

Allen designed his framework to support reasoning about intervals and it
comprises thirteen relations in total. 
The seven basic \emph{Allen's relations} are shown in the top half of Table~\ref{table:allen-relations}.
For example, interval $r$ \RelationName{meets} interval $s$ when $r$ finishes immediately before $s$ starts.
This is illustrated by the doodle in the table.
We will use smaller versions of the doodles also in the text: (\Doodle{\Meets}).
The first six relations in the table also have an inverse counterpart (hence thirteen relations).
For example, relation ``$r$ \RelationName{inverse meets} $s$'' describes $s$ immediately finishing before $r$ starts: (\Doodle{\Reverse{\Meets}}).

\begin{table}
  \caption{Allen's and ISEQL interval relations}
  \centering\small
  \AfterTableCaptionSpace
  \newcommand*{\join}{\JoinName}
  \begin{tabular}{@{}>{\raggedright}p{6em}p{20mm}p{12em}@{}}  
    \toprule
    Relation & \ Doodle & Formal definition \\
    \midrule
    \join{overlaps} & \Doodle[table]{\Overlaps} &$r.T_s < s.T_s < r.T_e < s.T_e$\\
    \addlinespace
    \join{during}   & \Doodle[table]{\During}   &$s.T_s < r.T_s \land r.T_e<s.T_e$\\
    \addlinespace
    \join{before}   & \Doodle[table]{\Before}   &$r.T_e < s.T_s$              \\
    \addlinespace
    \join{meets}    & \Doodle[table]{\Meets}    &$r.T_e = s.T_s$              \\
    \addlinespace
    \join{equals}   & \Doodle[table]{\Equals}   &$r.T_s = s.T_s \land r.T_e=s.T_e$\\
    \addlinespace
    \join{starts}   & \Doodle[table]{\Starts}   &$r.T_s = s.T_s \land r.T_e<s.T_e$\\
    \addlinespace
    \join{finishes} & \Doodle[table]{\Finishes} &$s.T_s < r.T_s \land
    r.T_e=s.T_e$\\
    \midrule  
    \join{start preceding} & \Doodle[table]{\StartPreceding}   &
    $r.T_s \le s.T_s < r.T_e $\newline
    $s.T_s - r.T_s \le \delta                $\\
    \addlinespace
    \join{end following}   & \Doodle[table]{\EndFollowing}     &
    $r.T_s < s.T_e \le r.T_e$\newline
    $r.T_e - s.T_e \le \varepsilon                $\\
    \addlinespace
    \join{before}          & \Doodle[table]{\Before}           &
    $r.T_e \le s.T_s$\newline
    $s.T_s - r.T_e \le \delta                $\\
    \addlinespace
    \join{left overlap}    & \Doodle[table]{\LeftOverlap}      &
    $r.T_s \le s.T_s < r.T_e \le s.T_e$\newline
    $s.T_s - r.T_s \le \delta                $\newline
    $s.T_e - r.T_e \le \varepsilon           $\\
    \addlinespace
    \join{during}          & \Doodle[table]{\During}           &
    $s.T_s \le r.T_s \land r.T_e \le s.T_e$ \newline
    $r.T_s - s.T_s \le \delta                $\newline
    $s.T_e - r.T_e \le \varepsilon           $\\
    \bottomrule
  \end{tabular}
  \label{table:allen-relations}
\end{table}


ISEQL originated in the context of complex event detection and covers a
different set of requirements. The list
of the five basic ISEQL relations is presented in the bottom half of
Table~\ref{table:allen-relations}; each of them has an inverse counterpart.
Additionally, ISEQL relations are parameterized.
The parameters control additional constraints and allow a much more
fine-grained definition of join predicates. This is similar to the simple
temporal problem (STP) formalism, which defines an interval that restricts the
temporal distance between two events~\cite{AFC13,DMP91}. 
Let us consider the \RelationName{before} ISEQL relation (\Doodle{\Before}).
It has one parameter $\delta$, which controls the maximum allowed distance between the intervals (events).
When $\delta = 0$, this relation is equivalent to the Allen's \RelationName{meets} (\Doodle{\Meets}).
When $\delta > 0$, it is a disjunction of Allen's \RelationName{meets} and \RelationName{before}, and the maximum allowed distance between the events is $\delta$ timepoints.
Any ISEQL relation parameter can be \emph{relaxed} (set to infinity), which removes the corresponding constraint.

\paragraph*{Joins on Interval Relations:\,} 

For each binary interval relation (Allen's or ISEQL) we define a predicate $P(r, s)$ as its indicator function: its value is `true' if the argument tuples satisfy the relation and `false' otherwise.
From now on we will use the terms `predicate' and `binary interval relation' interchangeably.
We perceive ISEQL relation parameters, if not relaxed, as part of the definition
of $P$ (e.g. $P_{\delta}$).
We also define a \emph{temporal relation} (not to be confused with binary relations described before) as a set of temporal tuples: $\Rel r = \{r_1, r_2, \dots, r_n\}$.

Let us take a generic relational predicate $P(r, s)$. We define the $P$-\emph{join} of two relations $\Rel r$ and $\Rel s$ as an operator that returns all pairs of $\Rel r$ and $\Rel s$ tuples that satisfy predicate $P$. We can express this in pseudo-SQL as ``\texttt{SELECT * FROM \Rel r, \Rel s WHERE $P(r, s)$}''. For example, we define the ISEQL \JoinName{left overlap join} as ``\texttt{SELECT * FROM \Rel r, \Rel s WHERE $r$ LEFT OVERLAP $s$}''. If the predicate is parameterized, the join operator will also be parameterized.

\begin{example}
As an example (see Figure~\ref{fig:example-relations}),
let us assume that we have two temporal relations: $\Rel r = \Set{
r_1 = [0, 1), r_2 = [1, 3), r_3 = [2, 5)}$ and $\Rel s = \Set{ s_1 = [1, 3),
s_2 = [3, 4)}$.

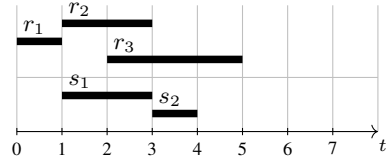
\begin{figure}[ht!]
\begin{tikzpicture}[relations, scale=0.6]
\DrawISEQLExampleRelations
\end{tikzpicture}
\caption{Example relations \Rel r and \Rel s}
\label{fig:example-relations}
\vspace*{-.2cm}
\end{figure}

Let us now take the ISEQL \JoinName{before join} (\Doodle{\Before}) with the parameter $\delta = 1$.
Its result consists of two pairs $\Tuple{r_1, s_1}$ and $\Tuple{r_2, s_2}$, because only they satisfy this particular join predicate.
If we relax the parameter (set $\delta$ to $\infty$), then an additional third pair $\Tuple{r_1, s_2}$ would be added to the result of the join.
\end{example}

By replacing the relational predicate by its formal
definition~(Table~\ref{table:allen-relations}),
we can implement an interval join in any relational database.
However, such an implementation results in a relational join with inequality
predicates, which
is not efficiently supported by RDBMSs: they have to fall back on
the nested-loop implementation in this case \cite{khayyat_fast_2017}.











\section{Formalizing our Approach}
\label{sec:formalization}

We opt for a relational algebra representation to be able to make formal
statements (e.g. proofs) about the different operators.  Before fully
formalizing our approach, we first introduce a map operator ($\chi$) as an
addition to the standard selection, projection, and join operators in
traditional relational algebra.  This operator is used for materializing
values as described in \cite{BMG93}.  We also introduce our new
interval-timestamp join ($\JoinByS$), allowing us to replace costly
non-equi-join predicates with an operator that, as we show later, can be
implemented much more efficiently.

\begin{definition}
The map operator $\Map{a}{e}(\Rel r)$ evaluates the expression $e$ on each
tuple of $\Rel r$ and concatenates the result to the tuple as attribute $a$:
\[
    \Map{a}{e}(\Rel r) = \SetBuilder{
        r \circ [a:e(r)]
    }{
        r \in \Rel r
    }.
\]
If the attribute $a$ already exists in a tuple, we instead overwrite its value.
\end{definition}

\begin{definition}
The interval-timestamp join $\Rel r \JoinByS \Rel s$ matches the intervals
in the tuples of relation $\Rel r$ with the timestamps of the tuples in $\Rel
s$. It comes in two flavors, depending on the timestamp chosen for $s$, i.e.,
$T_s$ or $T_e$. So, the interval-starting-timestamp join is defined as
\[
    \Rel r \JoinByTs^\theta \Rel s = \SetBuilder{
        r \times s
    }{
        r \in \Rel r, s \in \Rel s: r.T_s \;\theta\; s.T_s < r.T_e
    }
\]
with $\theta \in \{<, \leq\}$, whereas the interval-ending-timestamp join boils down to
\[
    \Rel r \JoinByTe^\theta \Rel s = \SetBuilder{
        r \times s
    }{
        r \in \Rel r, s \in \Rel s: r.T_s < s.T_e \;\theta\; r.T_e
    }
\]
with $\theta \in \{<, \leq\}$.

\end{definition}

Now we are ready to formulate the joins on interval relations shown in
Table~\ref{table:allen-relations} in
relational algebra extended by our new join operator $\JoinByS$. 
We first cover the non-parameterized versions (i.e., setting $\delta$ and
$\varepsilon$ to infinity) and then move on to the
parameterized ones. Table~\ref{table:mappingtora} gives an overview of the relational
algebra formulations. 
\iftechreport{
Proof for the correctness of our rewrites can be found in
Appendix~\ref{sec:rewrites}.
}
\ifnottechreport{
Proof for the correctness of our rewrites can be found in
Appendix~A of our technical report~\cite{ourtechreport}.
}

\begin{table*}
  \caption{Mapping of interval relations to relational algebra}
  \centering\small
  \AfterTableCaptionSpace
  \newcommand*{\join}{\JoinName}
  \begin{tabular}{llll}
    \toprule
    & Relation & \ Doodle & Relational algebra expression \\
    \midrule
    & \join{start preceding} & \Doodle[table]{\StartPreceding}   &
    $\Rel r \JoinByTssm^{\leq} \Rel s$ \\
    \addlinespace
    & \join{end following}   & \Doodle[table]{\EndFollowing}     &
    $\Rel r \JoinByTesm^{\leq} \Rel s$ \\
    \addlinespace
    \multirow{5}{*}{\rotatebox{90}{non-parameterized}} &
    \join{left overlap}    & \Doodle[table]{\LeftOverlap}      &
    $\Select_{\Rel r.T_e \leq \Rel s.T_e} (\Rel r \JoinByTssm^{\leq}  \Rel s)$ \\
    \addlinespace
    & \join{during}          & \Doodle[table]{\Reverse{\During}}           &
    \REV{$\Select_{\Rel r.T_e \leq \Rel s.T_e} (\Rel s \JoinByTssm^{\leq} \Rel r)$} \\
    \addlinespace
    & \join{before}          & \Doodle[table]{\Before}           &
    $\Map{T_e}{\infty}(\Map{T_s}{T_e}(\Rel r))\JoinByTssm^{\leq}\Rel s$ \REV{(ISEQL); $\Map{T_e}{\infty}(\Map{T_s}{T_e}(\Rel r))\JoinByTssm^{<}\Rel s$ (Allen)}\\
    \addlinespace
    & \join{meets}    & \Doodle[table]{\Meets}    &
    $\Map{T_e}{T_e + 1}(\Map{T_s}{T_e}( \Rel r ))\JoinByTssm^{\leq} \Rel s$ \\
    \addlinespace
    & \join{equals}   & \Doodle[table]{\Equals}   &
    $\Select_{\Rel r.T_e' = \Rel s.T_e} (\Map{T_e}{T_s + 1}(\Map{T_e'}{T_e}
    (\Rel r)) \JoinByTssm^{\leq} \Rel s)$ \\
    \addlinespace
    & \join{starts}   & \Doodle[table]{\Starts}   &
    $\Select_{\Rel r.T_e' < \Rel s.T_e} (\Map{T_e}{T_s + 1}(\Map{T_e'}{T_e}
    (\Rel r)) \JoinByTssm^{\leq} \Rel s)$ \\
    \addlinespace
    & \join{finishes} & \Doodle[table]{\Finishes} &
    $\Select_{\Rel s.T_s < \Rel r.T_s'} (\Map{T_s}{T_e - 1}(\Map{T_s'}{T_s}
    (\Rel r)) \JoinByTesm^{\leq} \Rel s)$ \\
    \midrule    
    & \join{start preceding} & \Doodle[table]{\StartPreceding}   &
    $\Map{T_e}{\min(T_e, T_s + \delta + 1)} (\Rel r) \JoinByTssm^{\leq} \Rel s$ \\
    \addlinespace
    \multirow{5}{*}{\rotatebox{90}{parameterized}} &
    \join{end following}   & \Doodle[table]{\EndFollowing}     &
    $\Map{T_s}{\max(T_s, T_e - \varepsilon - 1)} (\Rel r) \JoinByTesm^{\leq} \Rel s$ \\
    \addlinespace
    & \join{before}          & \Doodle[table]{\Before}           &
    $\Map{T_e}{T_e + \delta + 1}( \Map{T_s}{T_e}( \Rel r )) \JoinByTssm^{\leq} \Rel s$ \\
    \addlinespace
    & \join{left overlap}    & \Doodle[table]{\LeftOverlap}      &
    $\Select_{\Rel r.T'_e \leq \Rel s.T_e \leq \Rel r.T'_e + \varepsilon}
    (\Map{T_e}{\min(T_e, T_s + \delta + 1)} (\Map{T'_e}{T_e} (\Rel r)) \JoinByTssm^{\leq} \Rel s)$ \\
    \addlinespace
    & \join{during}          & \Doodle[table]{\Reverse{\During}}           &
    \REV{$\Select_{\Rel s.T'_e - \varepsilon \leq \Rel r.T_e \leq \Rel s.T'_e}
    (\Map{T_e}{\min(T_e, T_s + \delta + 1)} (\Map{T'_e}{T_e} (\Rel s)) \JoinByTssm^{\leq} \Rel r)$} \\
    \bottomrule
  \end{tabular}
  \label{table:mappingtora}
\end{table*}

\subsection{Non-parameterized Joins} 

The non-parameterized joins include all the Allen's relations and the ISEQL
joins with relaxed parameters. 
The \RelationName{equals}, \RelationName{starts}, \RelationName{finishes}, and
\RelationName{meets} predicates could be evaluated using a regular
equi-join. Nevertheless, we formulate them via interval-timestamp joins, which
are better suited to streaming environments and can be processed more
quickly for low  numbers of matching tuples. We present the joins
roughly in the order of their complexity, i.e., how many other
different operators we need to define them.

\paragraph*{ISEQL Start Preceding Join:\,} 
\label{sec:start-preceding-idea}

The \RelationName{start preceding}
predicate (\Doodle{\StartPreceding}) joins two tuples $r$ and $s$ if they
overlap and $r$ does not start after $s$ (we relax the
parameter $\delta$ here).
This can be expressed in our extended relational algebra in the following way:
\[
    \Rel r \JoinByTs^{\leq} \Rel s.
\]

%

\paragraph*{ISEQL End Following Join:\,}
\label{sec:end-following-idea}

As before, we first consider the ISEQL \JoinName{end following} predicate
(\Doodle{\EndFollowing}) with a relaxed $\varepsilon$ parameter, meaning that
tuples $r$ and $s$ should overlap and $r$ is not allowed to end before $s$. In
relational algebra this boils down to
\[
    \Rel r \JoinByTe^{\leq} \Rel s.
\]


\paragraph*{Overlap Join:\,}

If we look at the \RelationName{left overlap} join (\Doodle{\LeftOverlap}), we
notice that it looks very similar to a \RelationName{start preceding}
join. The main difference is that it has one additional constraint: the $\Rel
r$ tuple has to end before the $\Rel s$ tuple. Formulated in relational algebra this
is equal to 
\[
    \Select_{\Rel r.T_e \leq \Rel s.T_e} (\Rel r \JoinByTs^{\leq}  \Rel s).
\]


For the \RelationName{right overlap} join (\Doodle{\Reverse{\LeftOverlap}}),
we could just swap the roles of $\Rel r$ and $\Rel s$, or we could use an
\RelationName{end following} join combined with a selection predicate stating
that the $\Rel s$ tuple has to start before the $\Rel r$ tuple:
\[
    \Select_{\Rel s.T_s \leq \Rel r.T_s} (\Rel r \JoinByTe^{\leq}  \Rel s).
\]
For the more strict Allen's \RelationName{left overlap} and \RelationName{inverse overlap} joins
we use ``$<$'' for the $\theta$ of the join and the
selection predicate (or, alternatively, the parameterized version of the
\RelationName{overlap} join, which is introduced later).


\REV{
\paragraph*{During Join:\,}

For the \RelationName{during} join (\Doodle{\During}), we have to swap the
roles of $\Rel r$ and $\Rel s$. Formulated with the help of a
\RelationName{start preceding} join, it becomes
\[
    \Select_{\Rel r.T_e \leq \Rel s.T_e} (\Rel s \JoinByTssm^{\leq} \Rel r)
\]
or, alternatively, with an \RelationName{end following} join we get
\[
    \Select_{\Rel r.T_s \geq \Rel s.T_s} (\Rel s \JoinByTe^{\leq} \Rel r).
\]
A \RelationName{reverse during} join maps more naturally to a
\RelationName{start preceding} or \RelationName{end following} join, i.e., we
do not have to swap the roles of $\Rel r$ and $\Rel s$.
} For the Allen relation we use ``$<$'' for the $\theta$ of the join and the
selection predicate (or the parameterized version of the \RelationName{during} join).


\paragraph*{Before Join:\,} 
\label{sec:before-idea}

In the case of the \RelationName{before} predicate (\Doodle{\Before}), the
tuples should not overlap at all. We achieve this by converting the ending
events of $\Rel r$ into starting ones and setting the ending events to
infinity (see Figure~\ref{fig:allen-before-join}, the dashed lines are the
original tuples, the solid lines the newly created ones). Formulated in relational
algebra we get \REV{for the ISEQL version of \RelationName{before}:}
\[
    \Map{T_e}{\infty}(
        \Map{T_s}{T_e}(\Rel r)
    ) 
    \JoinByTs^{\leq} 
    \Rel s.
\]
For the Allen relation we use $\theta = $``$<$'' for the join.


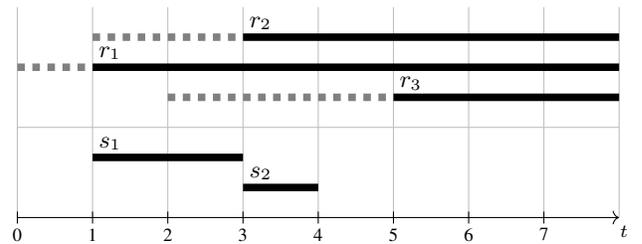
\begin{figure}[htb]
\centering
\begin{tikzpicture}[relations]
\renewcommand*{\DrawRTuple}[4]
{
    \DrawTuple[old tuple]{#1}{#2}{#3}{  }
    \DrawTuple           {#2}{8 }{#3}{#4}
}
\DrawISEQLExampleRelations
\ResetDrawRTuple
\end{tikzpicture}
\caption{Formulating Allen's \JoinName{before join}}
\label{fig:allen-before-join}
\end{figure}

\paragraph*{Meets Join:\,} 
\label{sec:meets-idea}

For the \RelationName{meets} predicate (\Doodle{\Meets}) each tuple of
relation $\Rel r$ should only be active for a short interval of length one
when it ends. We achieve this by converting the end events of tuples in $\Rel
r$ into start events and adding a new end event that shifts the old end event
by one. Expressed in relational algebra this looks as follows.
\[
    \Map{T_e}{T_e + 1}(\Map{T_s}{T_e}( \Rel r ))
    \JoinByTs^{\leq} 
    \Rel s.
\]


\paragraph*{Equals Join:\,} 

For the \RelationName{equals} predicate (\Doodle{\Equals}) 
we check that starting events match via an interval-time\-stamp join and
then add a selection to check the ending events:
\[
\Select_{\Rel r.T_e' = \Rel s.T_e} (\Map{T_e}{T_s + 1}(\Map{T_e'}{T_e}(\Rel
r)) \JoinByTs^{\leq} \Rel s).
\]

\paragraph*{Starts Join:\,} 

For a \RelationName{starts} predicate (\Doodle{\Starts})
we first check that the starting events are the same, the ending events
are again compared in a selection afterwards. Although one of the predicates uses a
comparison based on inequality, this happens in the selection, not the join:
\[
\Select_{\Rel r.T_e' < \Rel s.T_e} (\Map{T_e}{T_s + 1}(\Map{T_e'}{T_e}(\Rel
r)) \JoinByTs^{\leq} \Rel s).
\]

\paragraph*{Finishes Join:\,} 

The \RelationName{finishes} predicate (\Doodle{\Finishes}) works similar to a
\RelationName{starts} predicate. We use an interval-ending-timestamp join and
swap the roles of the starting and ending events (due to the different
definition of the $\JoinByTe^{\leq}$ join, we also have to shift
the timestamps by one):
\[
\Select_{\Rel s.T_s < \Rel r.T_s'} (\Map{T_s}{T_e - 1}(\Map{T_s'}{T_s}(\Rel
r)) \JoinByTe^{\leq} \Rel s).
\]

\subsection{Parameterized Joins} 

We now move on to the parameterized versions of the ISEQL join operators found
in Table~\ref{table:allen-relations}.

\paragraph*{Start Preceding Join with $\delta$:\,}

Defining a value for the parameter $\delta$ for the ISEQL \RelationName{start preceding} join (\Doodle{\StartPreceding}) means that the tuple from $\Rel s$
has to start between the start of the $\Rel r$ tuple and within $\delta$ time
units of the $\Rel r$ tuple starting or the end of the $\Rel r$ tuple
(whichever happens first). 
Basically, we shorten the long tuples.
Expressed in relational algebra, this becomes
\[
    \Map{T_e}{\min(T_e, T_s + \delta + 1)} (\Rel r) \JoinByTs^{\leq} \Rel s.
\]

%

\paragraph*{End Following Join with $\varepsilon$:\,}

For the parameterized \RelationName{end following} (\Doodle{\EndFollowing}) join we have to make sure
that the $\Rel s$ tuple ends within $\varepsilon$ time units distance from the end
of the $\Rel r$ tuple (but after the start of the $\Rel r$ tuple). The formal
definition in relational algebra is
\[
    \Map{T_s}{\max(T_s, T_e - \varepsilon - 1)} (\Rel r) 
    \JoinByTe^{\leq} 
    \Rel s.
\]

%

\paragraph*{The Before Join with $\delta$:\,}

For the parameterized \RelationName{before} join we have to make sure that the
$\Rel s$ tuple starts within a time window of length $\delta$ after the $\Rel
r$ tuple ends:
\begin{eqnarray*}
\Map{T_e}{T_e + \delta + 1}( \Map{T_s}{T_e}( \Rel r )) \JoinByTs^{\leq} \Rel s
\end{eqnarray*}


\paragraph*{Overlap Join with $\delta$ and $\varepsilon$:\,}

Similar to the non-parameterized version we use the
\RelationName{start preceding} join to define the
parameterized \RelationName{left overlap} join:
\[
    \Select_{\Rel r.T'_e \leq \Rel s.T_e \leq \Rel r.T'_e + \varepsilon}
        (\Map{T_e}{\min(T_e, T_s + \delta + 1)} (\Map{T'_e}{T_e} (\Rel r)) \JoinByTs^{\leq} \Rel s).
\]


For the parameterized \RelationName{right overlap} join we can either swap
the roles of $\Rel r$ and $\Rel s$ or use the 
\RelationName{end following} join:
\[
    \Select_{\Rel r.T'_s - \delta \leq \Rel s.T_s \leq \Rel r.T'_s}
        (\Map{T_s}{\max(T_s, T_e - \varepsilon - 1)} (\Map{T'_s}{T_s} (\Rel r)) \JoinByTe^{\leq} \Rel s).
\]


\REV{
\paragraph*{During Join with $\delta$ and $\varepsilon$:\,}

We can use a parameterized \RelationName{start preceding} or \RelationName{end
  following} join as a building block for a parameterized
\RelationName{during} join, swapping the roles of $\Rel r$ and $\Rel s$.
With a \RelationName{start preceding} join
we get
\[
    \Select_{\Rel s.T'_e - \varepsilon \leq \Rel r.T_e \leq \Rel s.T'_e}
    (\Map{T_e}{\min(T_e, T_s + \delta + 1)} (\Map{T'_e}{T_e} (\Rel s)) \JoinByTssm^{\leq} \Rel r),
\]


\noindent
whereas with an \RelationName{end following} join it boils down to
\[
    \Select_{
        \Rel s.T'_s \leq \Rel r.T_s \leq \Rel s.T'_s + \delta
    }(
        \Map{T_s}{\max(T_s, T_e - \varepsilon - 1)} (\Map{T'_s}{T_s} (\Rel s)) \JoinByTe^{\leq} \Rel r
    ).
\]

}

\section{Our Framework} 



After introducing the interval joins formally, we now turn to their efficient
implementation. We develop a framework to express the different interval joins
with the help of just one core join algorithm. The framework also includes an index
and several iterators for scanning through sets of intervals to increase the
performance and flexibility.

\subsection{The Endpoint Index} 
\label{sec:endpoint-index}

We can gain a lot of speed-up by sweeping through the
interval endpoints in chronological order using an Endpoint Index, 
which is a simplified version of the Timeline Index
\cite{kaufmann_timeline_2013}. The idea of the \emph{Endpoint Index} is that
intervals, which can be seen as points in a two-dimensional space, are mapped
onto one-dimensional \emph{endpoints} or \emph{events}.

Let $\mathbf r$ be an interval relation with tuples $r_i$, where $1\leq i\leq
n$. A tuple $r_i$ in an Endpoint Index is represented by two events of the form $e
= \Tuple{\mathit{timestamp},\allowbreak \mathit{type},\allowbreak tuple\_id}$,
where $\mathit{timestamp}$ is the $T_s$ or $T_e$ of the tuple, $\mathit{type}$
is either a $\mathrm{start}$ or $\mathrm{end}$ flag, and $\mathit{tuple\_id}$ is
the tuple identifier, i.e., the two events for a tuple $r_i$ are $\langle
r_i.T_s,\allowbreak \mathrm{start},\allowbreak i\rangle$ and $\langle
r_i.T_e,\allowbreak \mathrm{end},\allowbreak i\rangle$. For instance, for
$r_3.T = [3,\allowbreak 5)$, the two events are $\langle 3,\allowbreak
\mathrm{start},\allowbreak 3\rangle$ and $\langle 5,\allowbreak
\mathrm{end},\allowbreak 3\rangle$, which can be seen as ``at time 3 tuple 3
started'' and ``at time 5 tuple 3 ended''.

Since events represent timestamps, we can impose a total order among events, where the order is according to $\mathit{timestamp}$ and ties are broken by $\mathit{type}$.
In our case of half-open intervals, the order of $\mathit{type}$ values is: $\mathrm{end} < \mathrm{start}$.
Endpoints with equal timestamps and types but different tuple identifiers are considered equal.
An Endpoint Index for interval relation $\mathbf r$ is built by first
extracting the interval endpoints from the relation and then creating the
ordered list of events $\left[e_1, e_2, \dots, e_{2n}\right]$ sorted in
ascending order. In case of event detection, the endpoints (events) can be
taken directly from the event stream and we do not even have to construct an index.

Consider exemplary interval relation \Rel r from Figure~\ref{fig:example-relations}. The Endpoint Index for it is
$[\langle 0,\allowbreak \mathrm{start},\allowbreak 1\rangle$,
$ \langle 1,\allowbreak \mathrm{end},  \allowbreak 1\rangle$,
$ \langle 1,\allowbreak \mathrm{start},\allowbreak 2\rangle$,
$ \langle 2,\allowbreak \mathrm{start},\allowbreak 3\rangle$,
$ \langle 3,\allowbreak \mathrm{end},  \allowbreak 2\rangle$,
$ \langle 5,\allowbreak \mathrm{end},  \allowbreak 3\rangle]$.

%
%
%
%


\subsection{Endpoint Iterators} 
\label{sec:iterators}

Before continuing with the join algorithms, we introduce the concept of the Endpoint Iterator, upon which our family of algorithms is based. An \emph{Endpoint Iterator} represents a cursor, that allows forward traversing a list of endpoints (e.g., an Endpoint Index). More formally, it is an abstract data type (an interface), that supports three operations:

\begin{itemize}
\item \texttt{getEndpoint}: returns the endpoint, which the iterator is currently pointing to (initially returns the first endpoint in the list);
\item \texttt{moveToNextEndpoint}: advances the cursor to the next endpoint;
\item \texttt{isFinished}: return \texttt{true} if the cursor is pointing beyond the last endpoint of the list, \texttt{false} otherwise.
\end{itemize}

%

\noindent
More details on the implementation of Endpoint Iterators can be found in 
Appendix~\ref{appendix:iterators}.


The basic implementation of the Endpoint Iterator is the \emph{Index
  Iterator}, which provides an Endpoint Iterator interface to a physical
Endpoint Index. Given an instance of the index, such an iterator traverses
all Endpoint Index elements using the native method applicable to the
Endpoint Index. 
In the text and in the algorithm descriptions we use the terms ``Endpoint
Index'' and ``Index Iterator'' interchangeably, i.e., we create an Index
Iterator for an Endpoint Index implicitly if needed.

%
%
%
%
%
%


There are also \emph{wrapping} iterators. Such iterators do not have direct
access to an Endpoint Index, but modify, filter and/or combine the output of
one or several \emph{source} Endpoint Iterators. In software design pattern
terminology such iterators are called \emph{decorators}. We conclude this
subsection by introducing one such wrapping iterator. We introduce more of
them later, as needed.

The simplest wrapping Endpoint Iterator is the \emph{Filtering Iterator}. It
receives, upon construction, a source Endpoint Iterator and an endpoint type
($\mathrm{start}$ or $\mathrm{end}$). It then traverses only endpoints
having the specified type. 

%
%
%
%

\subsection{The Core Algorithm \texttt{JoinByS}} 
\label{sec:corealg}

We are now ready to define the core algorithm, which forms the basis of all our joins. This algorithm receives the relations to be joined \Rel r and \Rel s,
Endpoint Iterators for them, a comparison predicate, and a callback function
that will be called for each result pair. The algorithm performs the
interleaved scan of the endpoint iterators. While doing so, it maintains the
set of active \Rel r tuples. \emph{Every} endpoint for relation \Rel s
triggers the output---the Cartesian product of the corresponding tuple $s$ and
the set of active \Rel r tuples. The comparison predicate is used to define
the order in which equal endpoints of different relations are handled
(``equal'' meaning endpoints having the same timestamp and type).

The pseudocode for the core algorithm \emph{JoinByS} is presented in Algorithm~\ref{alg:join-by-s}.
The algorithm starts by initializing an active \Rel r tuple set implemented via a
map (an associative array) of tuple identifiers to tuples. 

\begin{algorithm2e}[htb]
    \caption{JoinByS(\Rel r, \Rel s, \textsf{itR}, \textsf{itS},
        \textsf{comp}, \textsf{consumer})}
    \label{alg:join-by-s}
    \KwData  {argument relations \Rel r and~\Rel s,
        corresponding Endpoint Iterators \ItR and~\ItS,
        endpoint comparison predicate \Comp (`$<$' or `$\leq$'),
        function \Consumer{$r, s$} for result pairs}
    \Var \ActiveR$ \leftarrow$ \New Map of tuple identifiers to tuples\;
    \While{\Not \ItR.\IsFinished \And \Not \ItS.\IsFinished}
    {
        \eIf{\Comp{\ItR.\GetEndpoint, \ItS.\GetEndpoint}}
        {
            \tcp{handle an \Rel r endpoint (maintain active \Rel r tuples)}
            $\mathit{tid} \leftarrow \ItR.\GetEndpoint.\mathit{tuple\_id}$\;
            \eIf{\ItR.\GetEndpoint$ = \mathrm{start}$}
            {
                $r \leftarrow \Rel r[\mathit{tid}]$\tcp*{load the tuple}
                \ActiveR.\Insert{$\mathit{tid}$, $r$}\;
            }
            {
                \ActiveR.\Remove{$\mathit{tid}$}\;
            }
            \ItR.\MoveToNextEndpoint\;
        }
        {
            \tcp{handle an \Rel s endpoint (trigger output)}
            $\mathit{s} \leftarrow \Rel s[\ItS.\GetEndpoint.\mathit{tuple\_id}]$\tcp*{load tuple $s$}
            \ForEach(\tcp*[f]{with every active tuple $r$}){r $\in$ \ActiveR}
            {
                \Consumer{$r$, $s$}\tcp*{produce output pair $\Tuple{r, s}$}
            }
            \ItS.\MoveToNextEndpoint\;
        }
    }
\end{algorithm2e}

The main loop (line 2) and the main ``if'' (line 3) implement the interleaved
scan of the endpoint indices (like in a sort-merge join). The tricky part here
is that instead of a hardwired comparison operator (`$<$' or `$\leq$'), we use
the function $\textsf{comp}$, that we pass as an argument to the algorithm.
In case of the \JoinName{start preceding join}, for instance, if both current
endpoints of \Rel r and \Rel s are equal, we have to handle the \Rel r
endpoint first (Section~\RefOnPage{sec:start-preceding-idea}), and thus we
have to use the `$\leq$' predicate. In case of the \JoinName{end following
  join}, on the other hand, if both current endpoints of \Rel r and \Rel s are
equal, we have to handle the \Rel s endpoint first
(Section~\RefOnPage{sec:end-following-idea}), and thus we have to use the
`$<$' predicate. Having the predicate as an argument of the algorithm allows
us to choose the needed predicate upon using the algorithm, which prevents
code duplication.\footnote{Note that the comparison function is not the
  same as the parameter $\theta$ of the interval-timestamp join. \REV{The
  comparison operator in JoinByS makes sure that the events are processed in
  the right order.}}

The rest of the algorithm consists of two parts. The first part (lines 4--10)
handles an \Rel r endpoint and manages the active \Rel r tuple set. When a
tuple starts, the algorithm loads it from the relation by the tuple identifier
stored in the endpoint and puts the tuple in the map using the
identifier as the key. When a tuple ends, the algorithm removes it from the
active tuple map, again using the tuple identifier as the key.

The second part (lines 12--15) handles an \Rel s endpoint. It first loads the
corresponding tuple $s$ from the relation. Then it iterates through all
elements in the active \Rel r tuple map. For every active $r$ tuple the
algorithm outputs the pair $\Tuple{r, s}$ by passing it into the
$\mathit{consumer}$ function, which is another function-type argument of the
algorithm. \REV{In some cases, the consumer has to do additional work such as
  evaluating a selection predicate. We call these consumers
  $\mathit{filteringConsumers}$. If they have access to the full tuple, they
  can check the predicate and immediately output a result tuple. In a
  streaming environment, we do not have access to the end events immediately,
  which means that a filteringConsumer also needs to buffer data until these
  events become available.}


\section{Assembling the Parts}

We now show how to construct the different interval relations using our
JoinByS operator and iterators. We start with the expressions from
Section~\ref{sec:formalization} that do not include map operators, followed by
those that do.

\subsection{Expressions Without Map Operators}


\paragraph*{Start Preceding and End Following Joins:\,}

These two join predicates are the easiest to implement, as they can be mapped
directly to the JoinByS operator. For the \JoinName{start preceding join} (\Doodle{\StartPreceding})
we have to keep track of the active
\Rel r tuples, and trigger the output by the \emph{start} of an \Rel s tuple. If
two tuples start at the same time, we have to handle the \Rel r tuple
first.
 Therefore, we call the JoinByS function, passing to it only the
starting \Rel s endpoints.  This is achieved by using a Filtering Iterator
(Section~\RefOnPage{sec:iterators}). We also have to pass the `$\leq$' predicate as
the comparison function.
A \JoinName{start preceding join} then boils down to a single call of JoinByS
(see Algorithm~\ref{alg:start-preceding-join-base}).

\begin{algorithm2e}
    \caption{StartPrecedingJoin(\Rel r, \Rel s,
        \textsf{itR}, \textsf{itS}, \textsf{consumer})}
    \label{alg:start-preceding-join-base}
    JoinByS(\Rel r, \Rel s, \ItR, FilteringIterator(\ItS, $\mathrm{start}$), `$\leq$', \Consumer)\;
\end{algorithm2e}

The algorithm StartPrecedingJoin receives iterators to the Endpoint Indexes. When using this algorithm with Endpoint Indices, we simply wrap each index in an Index Iterator---an operation, which, as noted before, we consider implicit.

We define the algorithm for the \JoinName{end following join} (\Doodle{\EndFollowing}) similarly, but filter the \emph{ending} endpoints of \Rel s, and pass the `$<$' as the comparison function. The pseudocode of the EndFollowingJoin is presented in Algorithm~\ref{alg:end-following-join-base}.

\begin{algorithm2e}
    \caption{EndFollowingJoin(\Rel r, \Rel s,
        \textsf{itR}, \textsf{itS}, \textsf{consumer})}
    \label{alg:end-following-join-base}
    JoinByS(\Rel r, \Rel s, \ItR, FilteringIterator(\ItS, $\mathrm{end}$), `$<$', \Consumer)\;
\end{algorithm2e}


\paragraph*{Overlap Joins:\,}

The \JoinName{left overlap join} (\Doodle{\LeftOverlap}) can be implemented using the StartPrecedingJoin
algorithm with an additional constraint $r.T_e \leq s.T_e$. The pseudocode is
shown in Algorithm~\ref{alg:left-overlap-join}. The 
\JoinName{right overlap join} (\Doodle{\Reverse\LeftOverlap}) 
is implemented along similar lines using the
EndFollowingJoin algorithm and the selection predicate
$s.T_s \leq r.T_s$.

\begin{algorithm2e}
    \caption{LeftOverlapJoin(\Rel r, \Rel s,
        \textsf{idxR}, \textsf{idxS}, \textsf{consumer})}
    \label{alg:left-overlap-join}
    \textsf{filteringConsumer} $\leftarrow$ \Function{$(r, s)$}
    {
        \lIf{$r.T_e \leq s.T_e$}
        {
            \Consumer{$r$, $s$}
        }
    }
    StartPrecedingJoin(\Rel r, \Rel s, \IdxR, \IdxS,
        \textsf{filteringConsumer})
\end{algorithm2e}

For the Allen versions of the overlap joins, we use strict versions of
Algorithms \ref{alg:start-preceding-join-base} and
\ref{alg:end-following-join-base}, \emph{StartPrecedingStrictJoin} and
\emph{EndFollowingStrictJoin}, which do not allow a tuple $r$ to start with a
tuple $s$ or a tuple $s$ to end with a tuple $r$, respectively. They are just
simple variations: StartPrecedingStrictJoin merely replaces the `$\leq$' in
Algorithm~\ref{alg:start-preceding-join-base} with `$<$' and
EndFollowingStrictJoin replaces the `$<$' in
Algorithm~\ref{alg:end-following-join-base} with `$\leq$'. 
Additionally, we change the `$\leq$' in the 
selection predicates in the
\textsf{filteringConsumer} functions to `$<$'.

\REV{

\paragraph*{During Joins:\,}

Implementing the \JoinName{during} join (\Doodle{\During}) is similar to
Algorithm~\ref{alg:left-overlap-join}: we just have to swap the arguments for
$\Rel r$ and $\Rel s$ (alternatively, we
could also use the \JoinName{end following} variant). For the Allen version of
\JoinName{during} joins, we replace the StartPrecedingJoin, EndFollowingJoin,
and selection predicates with their strict counterparts.

If we simply call an algorithm with
swapped arguments, the elements of the result pairs appear in a different
order, i.e., $\Tuple{s, r}$ instead of the expected $\Tuple{r, s}$. If this is
an issue, we can swap them back using a lambda function as the consumer.
Putting everything together, we get Algorithm~\ref{alg:during-join}.

\begin{algorithm2e}
    \caption{DuringJoin(\Rel r, \Rel s,
        \textsf{idxR}, \textsf{idxS}, \textsf{consumer})}
    \label{alg:during-join}
    \textsf{reversingConsumer} $\leftarrow$ \Function{$(s, r)$}
    {
        \Consumer{$r$, $s$}\;
    }
    \textsf{filteringConsumer} $\leftarrow$ \Function{$(s, r)$}
    {
        \lIf{$r.T_e \leq s.T_e$}
        {
            \textsf{reversingConsumer}($s$, $r$)
        }
    }
    StartPrecedingJoin(\Rel s, \Rel r, \IdxS, \IdxR,
        \textsf{filteringConsumer})
\end{algorithm2e}
}


\subsection{Expressions With Map Operators}

In order to avoid physically changing tuple values or even the Endpoint Index,
we apply the changes made by the map operators virtually with an iterator.
While performing an interleaved scan of two Endpoint Indexes, instead of simply comparing the two
endpoints $r^e$ and $s^e$ (as in $r^e < s^e$), we shift the timestamp of one
of them when comparing: $r^e + \delta < s^e$. In this way the algorithm performs an
interleaved scan of the indexes as if we had shifted all \Rel r tuples in time
by $+\delta$.

During an interleaved scan, instead of forcing the iterators of the two
Endpoint Indexes (for the relations \Rel r and \Rel s) to move synchronously
as in all the operators so far, now one of the iterators lags behind by a
constant offset. This
behavior can be easily incorporated into our framework by using a special
Endpoint Iterator that shifts the timestamp of every endpoint it returns on-the-fly.

There is a second issue: the new \emph{starting} endpoint often is actually a
shifted \emph{ending} endpoint or vice versa. Consequently, we have to change the endpoint
type as well. With the help of our \emph{Shifting Iterator}, we can shift timestamps
and also change endpoint types. As input parameters a shifting iterator
receives a source Endpoint Iterator, the shifting distance, and an endpoint
type (start or end).

%
%
%
%
%
%
%

The final issue is \REV{separately} shifting the starting and ending endpoints by
different amounts. We solve this by having independent iterators for both
starting and ending endpoints and merging them on-the-fly in an interleaved
fashion. The input parameters of the \emph{Merging Iterator} are two other
iterators, the events of which it merges. See Appendix \ref{appendix:iterators} for more details.

%
%
%


\paragraph*{Before and Meets Joins:\,}

We are now ready to create a
\emph{GeneralBeforeJoin} (see Algorithm~\ref{alg:before-join} and
\REV{Figure~\ref{fig:schematics} for a schematic representation}); we
already handle the parameterized version here as well. 
This algorithm performs a virtual three-way sort-merge join of the two
Endpoint Indexes. One pointer will traverse the Endpoint Index for relation
\Rel s, and two pointers will traverse the Endpoint Index for relation \Rel r,
all three pointers moving synchronously, but at different positions. This is
why we had to (implicitly) create two Index Iterators for the same index (lines 4 and 6)---each of
them represents a physical pointer to the same Endpoint Index, therefore we
need two of them.

\begin{algorithm2e}
    \SetInd{0pt}{1.5em}
    \caption{GeneralBeforeJoin(\Rel r, \Rel s,
        \textsf{idxR}, \textsf{idxS}, $\beta$, $\delta$, \textsf{consumer})}
    \label{alg:before-join}
    StartPrecedingJoin(\Rel r,\Rel s,\\\Indp
        MergingIterator(\\\Indp
            \tcp{$T_e + \beta \rightarrow T_s$}
            ShiftingIterator(\\\Indp
                FilteringIterator(\IdxR, $\mathrm{end}$),
                $\beta$, $\mathrm{start}$),\\\Indm
            \tcp{$T_e + \delta + 1\rightarrow T_e$}
            ShiftingIterator(\\\Indp
                FilteringIterator(\IdxR, $\mathrm{end}$),
                $\delta + 1$, $\mathrm{end}$)),\\\Indm\Indm
        IndexIterator(\IdxS),\\
        \Consumer)\;
\end{algorithm2e}

\REV{
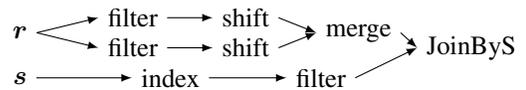
\begin{figure}[htb]
  \centering
  \begin{tikzpicture}
    \draw (0,0.6) node (A) {\Rel r};
    \draw (0,0) node (B) {\Rel s};
    \draw (1.5,0.8) node (C) {filter};
    \draw (1.5,0.4) node (D) {filter};
    \draw (2,0) node (E) {index};
    \draw (3,0.8) node (F) {shift};
    \draw (3,0.4) node (G) {shift};
    \draw (4.5,0.6) node (H) {merge};
    \draw (4,0) node (I) {filter};
    \draw (6,0.4) node (J) {JoinByS};
    \draw[-latex] (A.east) -- (C.west);
    \draw[-latex] (A.east) -- (D.west);
    \draw[-latex] (C.east) -- (F.west);
    \draw[-latex] (D.east) -- (G.west);
    \draw[-latex] (F.east) -- (H.west);
    \draw[-latex] (G.east) -- (H.west);
    \draw[-latex] (H.east) -- (J.west);
    \draw[-latex] (B.east) -- (E.west);
    \draw[-latex] (E.east) -- (I.west);
    \draw[-latex] (I.east) -- (J.west);
  \end{tikzpicture}
\caption{Schematic representation of \emph{GeneralBeforeJoin}}
\label{fig:schematics}
\end{figure}
}

We express the Allen's \JoinName{before join} (\Doodle{\Before}) by
substituting 1 and $+\infty$ for $\beta$ and $\delta$, respectively; Allen's
\JoinName{meets join} (\Doodle{\Meets}) by substituting 0 and 0,
respectively; and the ISEQL \JoinName{before join} by substituting 0 for
$\beta$ and only using the $\delta$ for the parameterized version. The
parameter $\beta$ distinguishes between the strict (Allen) and non-strict
(ISEQL) versions of the operator.


\paragraph*{Equals and Starts Joins:\,}

For the \JoinName{equals join} (\Doodle{\Equals}) we keep the original starting endpoints of \Rel
r and use as ending endpoints the starting endpoints shifted by one and then
execute a StartPrecedingJoin. This matches tuples from \Rel r and
\Rel s with the same starting endpoints. We check that we have matching ending
endpoints in the \textsf{filteringConsumer} function, which receives the
actual tuples as input and thus has access to the timestamp attributes of the
original tuples (see Algorithm~\ref{alg:equals-join}) for the pseudocode.

\begin{algorithm2e}
    \SetInd{0pt}{1.5em}
    \caption{EqualsJoin(\Rel r, \Rel s,
        \textsf{idxR}, \textsf{idxS}, \textsf{consumer})}
    \label{alg:equals-join}
    \textsf{filteringConsumer} $\leftarrow$ \Function{$(r, s)$}
    {
        \lIf{$r.T_e = s.T_e$}
        {
            \Consumer{$r$, $s$}
        }
    }
    StartPrecedingJoin(\Rel r,\Rel s,\\\Indp
        MergingIterator(\\\Indp
            \tcp{keep the original \Rel r starting endpoints}
                FilteringIterator(\IdxR, $\mathrm{start}$), \\
            \tcp{$T_s + 1\rightarrow T_e$}
            ShiftingIterator(\\\Indp
                FilteringIterator(\IdxR, $\mathrm{start}$),
                $1$, $\mathrm{end}$)),\\\Indm\Indm
        IndexIterator(\IdxS),\\
        \textsf{filteringConsumer})\;
\end{algorithm2e}

For a \JoinName{starts join} (\Doodle{\Starts}) we just have to change the predicate in the 
\textsf{filteringConsumer} function from `$=$' to `$<$'.


\paragraph*{Finishes Join:\,}

For the tuples in \Rel r we turn the ending events into starting events and
shift the ending events by one before joining them to the tuples in \Rel s via
an EndFollowingJoin (Algorithm~\ref{alg:end-following-join-base}). Finally, we
check that the tuple from \Rel s started before the one from \Rel r. For the
pseudocode of the \JoinName{finishes join} (\Doodle{\Finishes}), see
Algorithm~\ref{alg:finishes-join}.

\begin{algorithm2e}
    \SetInd{0pt}{1.5em}
    \caption{FinishesJoin(\Rel r, \Rel s,
        \textsf{idxR}, \textsf{idxS}, \textsf{consumer})}
    \label{alg:finishes-join}
    \textsf{filteringConsumer} $\leftarrow$ \Function{$(r, s)$}
    {
        \lIf{$s.T_s < r.T_s$}
        {
            \Consumer{$r$, $s$}
        }
    }
    EndFollowingJoin(\Rel r,\Rel s,\\\Indp
        MergingIterator(\\\Indp
            \tcp{\REV{$T_e - 1 \rightarrow T_s$}}
            ShiftingIterator(\\\Indp
                FilteringIterator(\IdxR, $\mathrm{end}$),
                \REV{$-1$}, $\mathrm{start}$),\\\Indm
            \tcp{\REV{$T_e \rightarrow T_e$}}
            ShiftingIterator(\\\Indp
                FilteringIterator(\IdxR, $\mathrm{end}$),
                \REV{$0$}, $\mathrm{end}$)),\\\Indm\Indm
        IndexIterator(\IdxS),\\
        \textsf{filteringConsumer})\;
\end{algorithm2e}


\paragraph*{Parameterized Start Preceding Join:\,}

We now turn to the parameterized variant of the \JoinName{start preceding join}
(\Doodle{\StartPreceding}), which has the parameter $\delta$ constraining the
maximum distance between tuple starting endpoints.
The basic idea is to take the starting endpoints of relation \Rel r, shift
them by $\delta + 1$, change their type to ending endpoints, and add these virtual
endpoints to the original endpoints of \Rel r.
This way each $r$ tuple will be represented by
three endpoints: the original starting and ending endpoints and the virtual ending
endpoint. Then the parameterless StartPrecedingJoin algorithm (Algorithm~\ref{alg:start-preceding-join-base}) is applied to
both streams of \Rel r and \Rel s endpoints. When encountering the second
ending endpoint in the merged iterator, it can simply be ignored when its
corresponding tuple cannot be found in the active tuple set (see Appendix~\ref{sec:firstend}).
Algorithm~\ref{alg:parametrized-start-preceding-join} depicts the pseudocode.

\begin{algorithm2e}
    \SetInd{0pt}{1.5em}
    \caption{PStartPrecedingJoin(\Rel r, \Rel s,
        \textsf{idxR}, \textsf{idxS}, $\delta$, \textsf{consumer})}
    \label{alg:parametrized-start-preceding-join}
    StartPrecedingJoin(\Rel r,\Rel s,\\\Indp
        \REV{FirstEndIterator}(\\\Indp
        MergingIterator(\\\Indp
            \tcp{keep the original \Rel r endpoints}
            IndexIterator(\IdxR),\\
            \tcp{$T_s + \delta + 1 \rightarrow T_e$}
            ShiftingIterator(\\\Indp
                FilteringIterator(\IdxR, $\mathrm{start}$),
                $\delta + 1$, $\mathrm{end}$))),\\\Indm\Indm
        IndexIterator(\IdxS),\\
        \Consumer)\;
\end{algorithm2e}

%
%
%
%


\paragraph*{Parameterized End Following Join:\,}

A similar parameterized \JoinName{end following join} 
(\Doodle{\EndFollowing}) is more
complicated. The problem here is that each $r$ tuple will have to be
represented by two starting endpoints. The algorithm must consider a
tuple activated only if \emph{both} starting endpoints (and no ending endpoint)
have been encountered.

We achieve this by introducing an iterator, called \emph{Second Start
  Iterator}, that stores the tuple identifiers of events for which we have only
encountered one starting endpoint in a hash set
\REV{(see Appendix~\ref{sec:secondstart})}.
Only the second starting
endpoint of this tuple will return the starting event. The pseudocode for the 
parameterized \JoinName{end following join} is shown in 
Algorithm~\ref{alg:parametrized-end-following-join}.

\begin{algorithm2e}
    \SetInd{0pt}{1.5em}
    \caption{PEndFollowingJoin(\Rel r, \Rel s,
        \textsf{idxR}, \textsf{idxS}, $\varepsilon$, \textsf{consumer})}
    \label{alg:parametrized-end-following-join}
    EndFollowingJoin(\Rel r,\Rel s,\\\Indp
        SecondStartIterator(\\\Indp
        MergingIterator(\\\Indp
            \tcp{keep the original \Rel r endpoints}
            IndexIterator(\IdxR),\\
            \tcp{$T_e - \varepsilon - 1 \rightarrow T_s$}
            ShiftingIterator(\\\Indp
                FilteringIterator(\IdxR, $\mathrm{end}$),
                $- \varepsilon - 1$, $\mathrm{start}$))),\\\Indm\Indm
        IndexIterator(\IdxS),\\
        \Consumer)\;
\end{algorithm2e}



\paragraph*{Parameterized Overlap Join:\,}

Now that we have an algorithm for the parameterized StartPrecedingJoin, we can
define the parameterized \JoinName{left overlap join} (\Doodle{\LeftOverlap}) by combining
PStartPrecedingJoin with a \textsf{filteringConsumer} function, similarly to what we
have done for the non-parameterized overlap join. 
Algorithm~\ref{alg:parameterized-left-overlap-join} shows the pseudocode.
Alternatively, we can use a PEndFollowingJoin and then check the predicate for
the starting endpoint of the $s$ tuple in the \textsf{filteringConsumer} function.

\begin{algorithm2e}
    \caption{PLeftOverlapJoin(\Rel r, \Rel s,
        \textsf{idxR}, \textsf{idxS}, $\delta$, $\varepsilon$, \textsf{consumer})}
    \label{alg:parameterized-left-overlap-join}
    \textsf{filteringConsumer} $\leftarrow$ \Function{$(r, s)$}
    {
        \lIf{$r.T_e \leq s.T_e \leq r.T_e + \varepsilon$}
        {
            \Consumer{$r$, $s$}
        }
    }
    PStartPrecedingJoin(\Rel r, \Rel s, \IdxR, \IdxS, $\delta$,
        \textsf{filteringConsumer})
\end{algorithm2e}

The \JoinName{right overlap join} (\Doodle{\Reverse\LeftOverlap}) uses a
PEndFollowingJoin with the corresponding predicate in the
\textsf{filteringConsumer} function.

\REV{

\paragraph*{Parameterized During Join:\,} 

The parameterized \JoinName{during join} (\Doodle{\During}) looks similar to
Algorithm~\ref{alg:parameterized-left-overlap-join}, we apply changes along
the lines of those shown in the paragraph for the non-parameterized
\JoinName{during join}. (There is also an alternative
version using an PEndFollowingJoin.)
}

\subsection{Correctness of Algorithms}
\label{sec:correctness}

Showing the correctness of our algorithms boils down to illustrating that we handle
the map operators correctly and demonstrating the
correctness of the StartPreceding and EndFollowing joins, as our algorithms
are either StartPreceding and EndFollowing joins or are built on top them.


\paragraph*{Iterators and Map Operators:\,}

Here we show how to implement map operators with the help of iterators.
Instead of materializing the result (e.g.\ on disk), we
make the corresponding changes in a tuple as it passes through an iterator. If
we still need a copy of the old event later on, we feed this event through
another iterator and merge the two tuple streams using a merge
iterator.


\paragraph*{StartPreceding Join:\,}

We have to show that all tuples created by
Algorithm~\ref{alg:start-preceding-join-base} satisfy the predicate
$r.T_s \le s.T_s < r.T_e$.  A Filter Iterator removes all the ending events
from \Rel s, so we only have to deal with starting events from \Rel s and with
both types of events from \Rel r. As comparison operator we use `$\le$'. This
determines the order in which events are dealt with.

First, let us look at the case that both upcoming events in \textsf{itR}
and \textsf{itS} are starting events. If $r.T_s \le s.T_s$, then $r$ will be
inserted into the active tuple set before $s$ is processed, meaning that the
(later) arrival of $s$ will trigger the join with $r$. If $r.T_s > s.T_s$,
then $s$ will be processed first, not encountering $r$ in the active tuple
set, meaning that the two will not join.

Second, if the next event in \Rel r is an ending event and the next event
in \Rel s a starting event, then the two events can never be equal. Even if
they have the same timestamp, the ending endpoint of $r$ will always be
considered less than the starting endpoint of $s$. Therefore, if $r.T_e \le
s.T_s$, $r$ will be removed first, so $r$ and $s$ will not join, and if $r.T_e
> s.T_s$, $s$ will still join with $r$. 

So, in summary, all the tuples generated by
Algorithm~\ref{alg:start-preceding-join-base} satisfy the predicate $r.T_s \le
s.T_s < r.T_e$.

For a StrictStartPreceding join we run
Algorithm~\ref{alg:start-preceding-join-base} with the comparison operator
`$<$', yielding output tuples that satisfy the predicate
$r.T_s < s.T_s < r.T_e$. If both upcoming events in \textsf{itR}
and \textsf{itS} are starting events, we get the correct behavior: 
$r.T_s < s.T_s$ will lead to a join, $r.T_s \ge s.Ts$ will not.
If the $r$ event is an ending event and the $s$ event is a starting one, we
also get the correct behavior: $r.T_e \le s.T_s$ will not join the $r$ and $s$
tuple, $r.T_e > s.T_s$ will (the ending event of $r$ is always less than the
starting event of $s$).


\paragraph*{EndFollowing Join:\,}

We show that all tuples created by Algorithm~\ref{alg:end-following-join-base}
satisfy the predicate $r.T_s < s.T_e \le r.T_e$.  This time a Filter Iterator
removes all the starting events from \Rel s, so we only have to deal with
ending events from \Rel s and with both types of events from \Rel r. The
comparison operator used for the non-strict version is `$<$'.

First, assume that the next event in \textsf{itR} is a starting event and the
next event in \textsf{itS} is an ending event. As an ending event takes
precedence over a starting event, if $r.T_s = s.T_e$, the $s$ event will come
first. In turn this means that if $r.T_s < s.T_e$, $r$ is added to the active
set first, resulting in a join, and if $r.T_s \ge s.T_e$, $s$ is processed
first, meaning there is no join.

Second, we now look at the case that both events are ending events. Due to the
comparison operator `$<$', the events are handled in the right way: if $r.T_e
< s.T_e$, we remove $r$ first, so there is no join, and if $r.T_e \ge s.T_e$
we handle $s$ first, resulting in a join.

For a StrictEndFollowing join we run
Algorithm~\ref{alg:end-following-join-base} with `$\le$' as comparison operator
to obtain tuples that satisfy the predicate $r.T_s < s.T_e < r.T_e$. Let us
first look at a starting event for \Rel r and an ending event for \Rel s. As
ending events are processed before starting events with the same timestamp, we
get: if $r.T_s < s.T_e$, then $r$ is added first, resulting in a join, and if
$r.T_s \ge s.T_e$, then $s$ is removed first, meaning there is no
join. Finally, we investigate the case that both events are ending events: if
$r.T_e \le s.T_e$, then $r$ is removed first, i.e., no join, and if $r.T_e >
s.T_e$, then $s$ is processed first, joining $r$ and $s$.

\section{Implementation Considerations} 

In this section we look at techniques to implement our framework efficiently,
in particular how to represent an active tuple set, utilizing contemporary hardware.
We also investigate the overhead caused by our heavy use of
abstractions (such as iterators).

\subsection{Managing the Active Tuple Set}
\label{sec:active-tuple-maps}

For managing the active tuple set we need a data structure into which we can
{\tt insert} key-value pairs, {\tt remove} them, and quickly enumerate (scan)
one by one all the values contained in the data structure via the
operation {\tt getnext}. In our case, the keys are
tuple identifiers and the values are the tuples themselves. The data structure
of choice here is a map or associative array.

The most efficient implementation of a map optimizing the {\tt insert} and
{\tt remove} operations is a hash table (with $O(1)$ time complexities for
these operations). However, hash tables are not well-suited for scanning.
The \emph{std::unordered\_map} class in the C++ Standard Template Library
and the \emph{java.util.HashMap} in the
Java Class Library, for instance, scan through all the buckets of a hash
table, making the performance of a scan operation linear with respect to the
capacity of the hash table and not to the actual amount of elements in it.

In order to achieve an $O(1)$ complexity for {\tt getnext}, the elements in
the hash table can be connected via a doubly-linked list (see
Figure~\ref{fig:linked-hash-map}).
The hash table stores pointers to elements, which in turn contain a key,
a value, two pointers for the doubly-linked list
(\emph{list prev} and \emph{list next}) and a pointer for chaining elements of
the same bucket for collision resolution (pointer \emph{bucket next}).
This approach is employed in the \emph{java.util.LinkedHashMap} in the Java Class
Library.

\begin{figure}[htb]
\centering
\includegraphics[width=6.1cm]{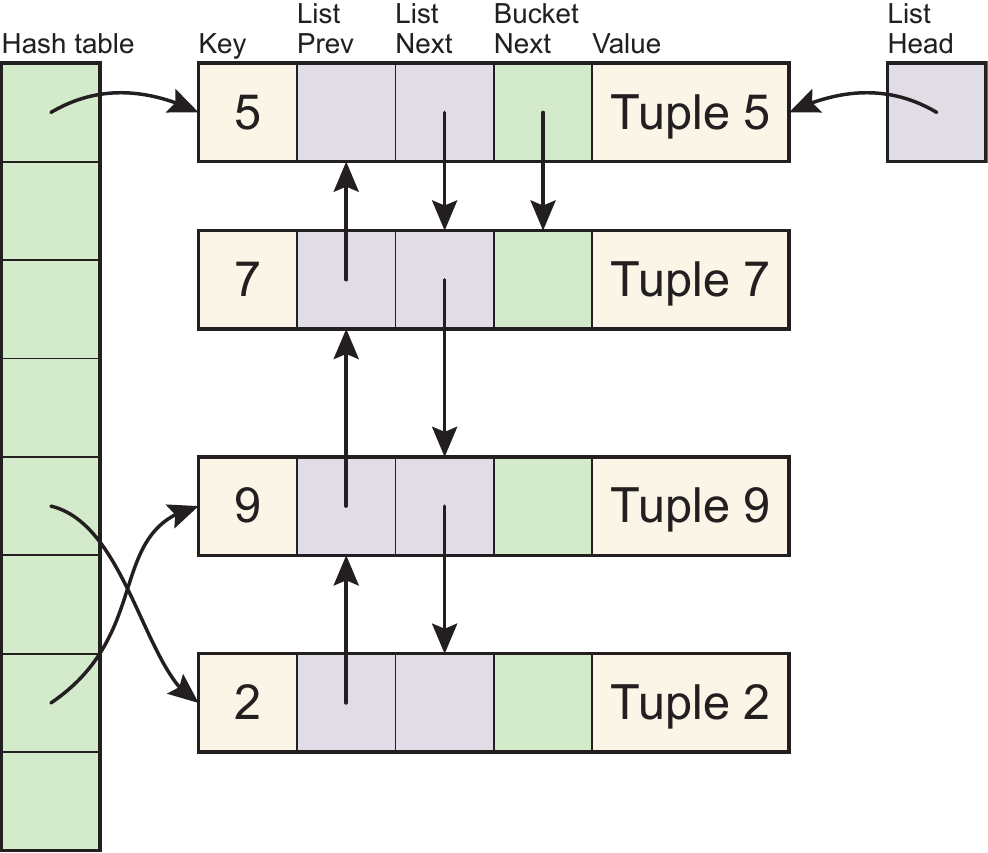}
\vspace*{-.15cm}
\caption{Linked hash map}
\label{fig:linked-hash-map}
\end{figure}

While this data structure offers a constant complexity for {\tt getnext}, the
execution times of different calls of {\tt getnext} can vary widely in
practice, depending on the memory footprint of the map. After a series of
insertions and deletions the elements of the linked list
become randomly scattered in memory, which \REV{has an impact on caching:
  sometimes the next list element is still in the cache (resulting in fast
  retrieval), sometimes it is not (resulting in slow random memory
  accesses). Additionally, the pointer structure make it hard for a prefetcher to
  determine where the next elements are located.}
However, for our approach it is crucial that
{\tt getnext} can be executed very efficiently, as it is typically called
much more often than \texttt{insert} and \texttt{remove}.
We will see in Section~\ref{sec:map} how to implement a hash map more
efficiently.

\subsection{Lazy Joining of the Active Tuple Set}
\label{sec:lazyjoining}

The fastest {\tt getnext} operations are actually those that are not executed.
We modify our algorithm to boost its performance by significantly
reducing the number of {\tt getnext} operations needed to generate the output.

We illustrate our point using the example setting in
Figure~\ref{fig:example-data}. Assume we have just encountered the left
endpoint of $s_1$, which means that our algorithm now scans the tuple set
$active^r$, which contains $r_1$ and $r_2$. After that we scan it again and
again when encountering the left endpoints of $s_2$, $s_3$, and
$s_4$. However, since no endpoints of $\mathbf r$ were encountered during that
time, we scan the same version of $active^r$ four times. We can reduce this to
one scan if we keep track of the tuples $s_1$, $s_2$, $s_3$, and $s_4$ in a
(contiguous) buffer, delaying the scan until there is about to be a change in
$active^r$.

\begin{figure}[htb]
\includegraphics[width=\linewidth]{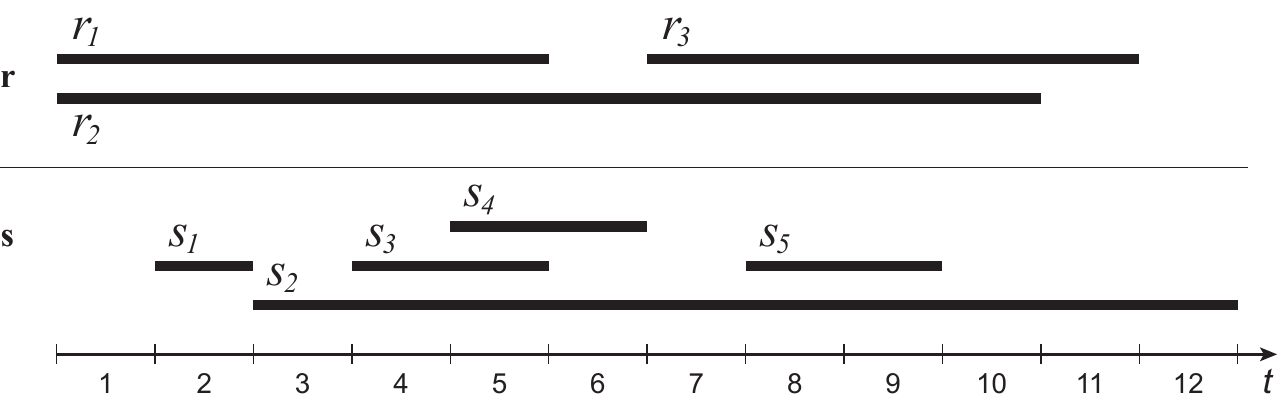}
\vspace*{-.15cm}
\caption{Example interval relations}
\label{fig:example-data}
\end{figure}

To remedy this situation, we collect all consecutively encountered \Rel s
tuples in a small buffer that fits into the L1 cache.
Scanning the active tuple set when producing the output now requires only one
traversal. 
Thanks to the design of our join algorithms we can incorporate this
optimization into the whole framework by modifying JoinByS. The optimized
version is shown in Algorithm~\ref{alg:lazy-join-by-s}. This technique has
been introduced for overlap joins in~\cite{piatov_interval_2016}, here
we generalize it to the JoinByS algorithm. We recommend using a
size for the buffer $c$ that is smaller than the size of the L1d CPU cache
(usually 32 Kilobytes) for this method to be effective.

%
%
\begin{algorithm2e}
    \caption{LazyJoinByS(\Rel r, \Rel s, \textsf{itR}, \textsf{itS},
        \textsf{comp}, \textsf{consumer})}
    \label{alg:lazy-join-by-s}
    \KwData  {argument relations \Rel r and~\Rel s,
        corresponding Endpoint Iterators \ItR and~\ItS,
        endpoint comparison predicate \Comp (`$<$' or `$\leq$'),
        function \Consumer{$r, s$} for result pairs}
    \Var \ActiveR$ \leftarrow$ \New Map of tuple identifiers to tuples\;
    \Var \Buffer$ \leftarrow$ \New array of capacity $c$\;
    \While{\Not \ItR.\IsFinished \And \Not \ItS.\IsFinished}
    {
        \eIf{\Comp{\ItR.\GetEndpoint, \ItS.\GetEndpoint}}
        {
            \tcp{handle an \Rel r endpoint (maintain active \Rel r tuples)}
            $\mathit{tid} \leftarrow \ItR.\GetEndpoint.\mathit{tuple\_id}$\;
            \eIf{\ItR.\GetEndpoint$ = \mathrm{start}$}
            {
                $r \leftarrow \Rel r[\mathit{tid}]$\tcp*{load the tuple}
                \ActiveR.\Insert{$\mathit{tid}$, $r$}\;
            }
            {
                \ActiveR.\Remove{$\mathit{tid}$}\;
            }
            \ItR.\MoveToNextEndpoint\;
        }
        {
            \tcp{get sequence of $s$ tuples uninterrupted by $r$ events}
            \Repeat{\ItS.\IsFinished \Or \Comp{\ItR.\GetEndpoint, \ItS.\GetEndpoint} \Or \Buffer.\IsFull}
            {
                $\mathit{s} \leftarrow \Rel s[\ItS.\GetEndpoint.\mathit{tuple\_id}]$\;
                \Buffer.\Insert{$s$}\;
                \ItS.\MoveToNextEndpoint\;
            }
            \tcp{produce Cartesian product with active \Rel r tuples}
            \ForEach(\tcp*[f]{scan the active $r$ tuples once}){r $\in$ \ActiveR}
            {
                \ForEach(\tcp*[f]{the inner loop, in L1 cache}){s $\in$ \Buffer}
                {
                    \Consumer{$r$, $s$}\tcp*{produce output pair}
                }
            }
            \Buffer.\Clear\;
        }
    }
\end{algorithm2e}

For the sake of simplicity, we only refer to the JoinByS algorithm in the
following section. It can be replaced by the LazyJoinByS algorithm without any
change in functionality.

\subsection{Features of Contemporary Hardware}

Before describing further optimizations, we briefly review mechanisms employed by
contemporary hardware to decrease main memory latency. This latency can have a
huge impact, as fetching data from main memory may easily use up more than a
hundred CPU cycles.

\paragraph*{Mechanisms:\,}

Usually, there is a hierarchy of caches, with smaller, faster ones closer to
CPU registers. Cache memory has a far lower latency than main memory, so a CPU
first checks whether the requested data is already in one of the caches
(starting with the L1 cache, working down the hierarchy). Not finding data in
a cache is called a {\em cache miss} and only in the case of cache misses on
all levels, main memory is accessed.
In practice an algorithm with a small memory footprint
runs much quicker, because in the ideal case, when an algorithm's data (and
code) fits into the cache, the main memory only has to be accessed
once at the very beginning, loading the data (and code) into the cache.

Besides the size of a memory footprint, the access pattern also plays a
crucial role, as contemporary hardware
contains {\em prefetchers} that speculate on which blocks of memory will be
needed next and preemptively load them into the cache. The easier the access
pattern can be recognized by a prefetcher, the more effective it
becomes. Sequential access is a pattern that can be picked up by prefetchers
very easily, while random access effectively renders them useless.

Also, programs do not access physical memory directly, but
through a virtual memory manager, i.e., virtual addresses have to be mapped to
physical ones. Part of the mapping table is cached in a so-called {\em
  translation lookaside buffer} (TLB).
As the size of the TLB is limited, a program with a high level of locality
will run faster, as all look-ups can be served by the TLB.

{Out-of-order execution} (also called {\em dynamic execution}) allows a CPU to
deviate from the original order of the instructions and run them as the data they
process becomes available. Clearly, this can only be done when the
instructions are independent of each other and can be run concurrently
without changing the program logic.

Finally, certain properties of DRAM (dynamic random access memory)
chips also influence latency. Accessing memory using fast
page or a similar mode means accessing data stored within the same page
or bank without incurring the overhead of selecting it. This mechanism
favors memory accesses with a high level of locality.

\paragraph*{Performance Numbers:\,}

We provide some numbers to give an impression of the performance of currently
used hardware.
For contemporary processors, such as ``Core'' and ``Xeon'' by
Intel\footnote{We use the cache and memory latencies
obtained for the Sandy Bridge family of Intel CPUs using the SiSoftware Sandra
benchmark, \url{http://www.sisoftware.net/?d=qa\&f=ben\_mem\_latency}.},
one random memory access within the L1 data (\emph{L1d}) cache (32~KB per
core) takes 4 CPU cycles. Within the L2
cache (256~KB per core) one random memory access takes 11--12 cycles. Within
the L3
cache (3--45~MB) one random memory access takes 30--40 CPU cycles. Finally, one
random physical RAM access takes around 70--100~ns (200--300 processor
cycles). It follows that the performance gap between an L1 cache access and a
main memory access is huge: two orders of magnitude.

\subsection{Implementation of the Active Tuple Set} 
\label{sec:map}

As we will see later in an experimental evaluation, managing the active tuple
set efficiently in terms of memory accesses is crucial for the performance of
the join algorithm. Otherwise we run the risk of starving the CPU while
processing a join. Our goals have to be to store the active tuple set as
compactly as possible and to access it sequentially, allowing the hardware to
get the data to the CPU in an efficient manner.

We store the elements of our hash map in a contiguous memory area. For the
{\tt insert} operation this means that we always append a new element at the
end of the storage area. Removing the last element from the storage area is
straightforward. If the element to be removed is not the last
in the storage area, we swap it with the last element and then remove it.
When doing so, we have to update all the references to the swapped elements.
Scanning involves stepping through the contiguous storage area sequentially.
We call our data structure a \emph{gapless hash map} (see
Figure~\ref{fig:gapless-hash-map}).

\begin{figure}[htb]
\centering
\includegraphics[width=6.1cm]{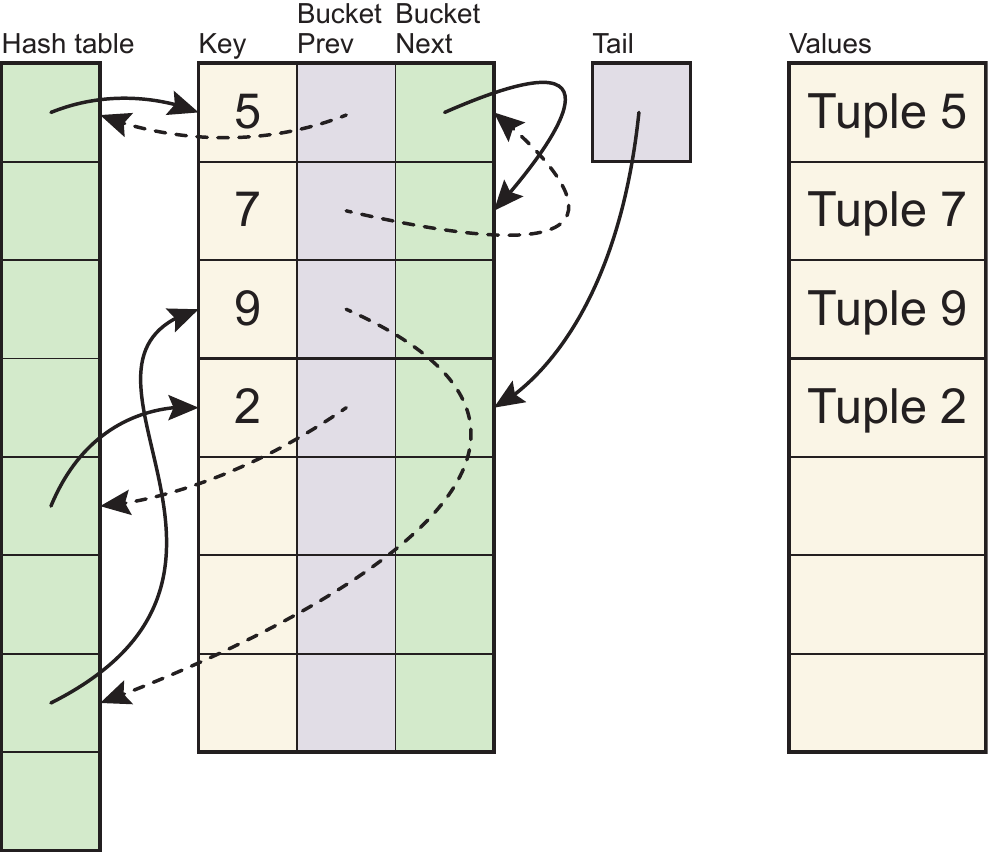}
\vspace*{-.15cm}
\caption{Gapless hash map}
\label{fig:gapless-hash-map}
\end{figure}

We also separate the tuples from the elements,
storing them in a different contiguous memory area in corresponding locations.
\REV{Assuming fixed-size records}, all basic element operations (append and
move) are mirrored for the corresponding tuples.
This slightly increases the costs for insertions and removal of
tuples. However, scanning the tuples is as fast as it can become, because we
do not need to read any metadata, only tuple information.

The hash table stores pointers to elements, which contain a key,
a pointer for chaining elements of the same
bucket when resolving collisions (pointer \emph{bucket next}, solid arrows),
and a pointer \emph{bucket prev} to a hash table entry or an element (whichever
holds the forward pointer to this element, dashed arrows). The latter is used
for updating the reference to an element when changing the element position. The
main difference to the random memory access of a linked hash map
(Fig.~\ref{fig:linked-hash-map}) is the allocation of all elements in a
contiguous memory area, allowing for fast sequential memory access when
enumerating the values.

\begin{example}
Assume we want to remove tuple 7 from the structure depicted in
Figure~\ref{fig:gapless-hash-map}. First of all, the bucket-next pointer of
the element with key 5 is set to NULL. Next, the last element in the storage
area (tuple 2) is moved to the position of the element with key 7. Following
the bucket-prev pointer of the just moved element we find the reference to the
element in the hash table and update it. Finally, the variable \emph{tail} is
decremented to point to the element with key 9.
\end{example}


%
%
%

\subsection{Overhead for Abstractions} 

All the abstractions we use (iterators, predicates passed as function
arguments, and lambda functions) allow us to express all joins by means of a
single function, which is extremely practical due to the huge simplification
of implementation and subsequent maintenance of the code. In this section we
explain why the impact of this architecture on the performance is minimal for
C++ and not significant for Java.

We  compare our implementation empirically to a manual
rewrite without abstractions of a selected join algorithm.
Here, we show the results for our most complicated implementation,
Algorithm~\ref{alg:before-join}. We compare its performance 
to a version that was fully inlined manually into a single leaf function. We
did so for C++ and also for Java. We then launched each one of the four
versions separately using the synthetic dataset of $10^6$ tuples with an
average number of active tuples equal to $10$ (see Section~\ref{sec:setup} for the dataset). Each version was executing the join 
several times sequentially to allow the JVM to perform all necessary
optimizations. The results are shown in
Figure~\ref{fig:experiments-compilers}. We see that the C++ version is several
times faster than the Java version. Moreover, we see that the C++ compiler was
able to optimize our abstracted code so well that its performance is
indistinguishable from the manually optimized version. The situation with Java
is more complicated, in the end the manually optimized version was
${\sim}10\%$ faster.

\begin{figure}[t]
\centering
\begin{tikzpicture}[baseline]
\ReadTable{normal-vs-inlined.txt}
\begin{axis}
[
    global plot style,
	xlabel = {Iteration sequential number},
	ylabel = {Execution time, s},
    xmin = 1, xmax = 13, xtick = data,
    ymin = 0, ymax = 1,
    legend columns = 2,
    height = 2.8cm,
    width = 7cm,
]
\AddPlot[]{x = run, y = java-n} \addlegendentry{Java normal}
\AddPlot[]{x = run, y = java-i} \addlegendentry{Java inlined}
\AddPlot[]{x = run, y = cpp-n}  \addlegendentry{C++ normal}
\AddPlot[]{x = run, y = cpp-i}  \addlegendentry{C++ inlined}
\end{axis}
\end{tikzpicture}
\caption{Overhead of the abstractions used in the algorithms}
\label{fig:experiments-compilers}
\end{figure}
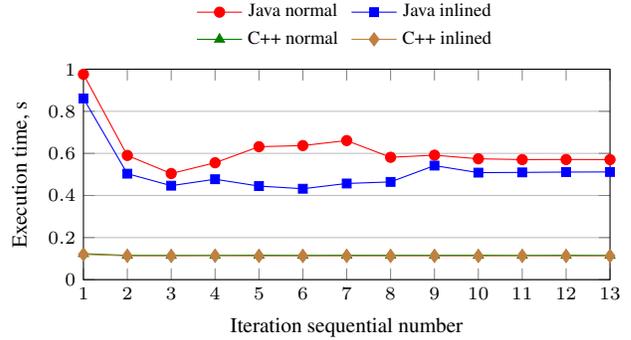

\paragraph*{C++:\,}

This language was designed to support penalty-free abstractions. Not
all abstractions in C++ are penalty-free, though.
We first implemented the family of Endpoint Iterators as a hierarchy of
virtual classes and found that the compilers we used (GCC and Clang) were not
able to inline virtual method calls (even though they had all the required information to do so). We then rewrote the code using templates
and functors, each iterator becoming a non-virtual class, passed into the join
algorithm as template argument. The
comparator used by the core algorithm was a functor \texttt{std::less} or
\texttt{std::less\_equals}. The consumers were defined as C++11
lambda-functions, also passed as a template parameter.
This time both compilers were able to inline all method calls and
generate very optimized code with all variables (including iterator fields)
kept in CPU registers.

\paragraph*{Java:\,}

We face a different situation with Java, as the optimization is not performed
by the compiler, but the Java virtual machine (JVM) during run-time. The JVM
(in particular, the standard Oracle HotSpot implementation) compiles,
optimizes and recompiles the code while executing it. It can potentially apply
a wider range of optimizations (e.g., speculative optimization) than a C++
compiler can, as it actively learns about the actual workload, but in the case
of Java we have limited control over this process.  As we show in
Figure~\ref{fig:experiments-compilers}, Java does in fact optimize the code
with abstractions. Not as well as C++, but the performance difference is very
small compared to a manually rewritten join.

\REV{
\subsection{Parallel Execution} 

While parallelization is not a main focus of this paper, we know how to
parallelize our scheme and have implemented a parallel version of our earlier
EBI-Join operator~\cite{piatov_interval_2016}. We give a brief description
here: the tuples in both input relations, \Rel r and \Rel s, are sorted by
their starting time and then partitioned in a round-robin fashion, i.e., the
$i$-th tuple of a relation is assigned to partition $(i \mod k)$ of that
relation, where $k$ is the number of partitions. By assigning close neighbors
to different partitions, we lower the size of the active tuple sets, which is
a crucial parameter for the performance of our algorithm. We then do a
pairwise join between all partitions of \Rel r with all partitions of \Rel
s. As all partitions are disjoint, the joins can run in parallel independently
of each other. A downside of this approach is that we need $k^2$
processes. Nevertheless, we achieved an average speed-up of 2.7, 4.3, and 5.3
for $k = $ 2, 3, and 4, respectively, on a machine with two CPUs (eight cores
each).

One major difference between JoinByS and EBI-Join is that JoinByS maintains
only one active tuple set (for \Rel r), whereas EBI-Join maintains two (one
for \Rel r and one for \Rel s). So, in order to keep the active tuple set
small, for JoinByS we only need to partition \Rel r, resulting in one process
for each of the $k$ partitions. The tuples in \Rel s are fed to each of these processes. 
}

\section{Theoretical Analysis} 
\label{sec:theoretical}


\paragraph*{\REV{One-dimensional Overlap:\,}}

Our approach is related to finding all the intersecting line segments, or
intervals, given a set of $n$ segments in one-dimensional space. The optimal
(plane-sweeping) algorithm for doing so has complexity $\BigO(n \log n + k)$,
where $k$ is the number of intersecting segments~\cite{Cormen09}.
\REV{In the worst case, when we have a large number of intersecting segments, the
complexity becomes $\BigO(n \log n + n^2)$. In this case, the run time of the
algorithm is dominated by the output cardinality.}

Each segment $s_i$ is split
up into a left starting event $\langle l_i, \mathrm{start} \rangle$ and a
right ending event $\langle r_i, \mathrm{end} \rangle$. Afterwards the events of all
segments are sorted, which takes $\BigO(n \log n)$ time. We then traverse the
sorted list of events. When encountering a left endpoint, we insert it into a
data structure $D$, which keeps track of the currently active segments. When
encountering a right endpoint, we remove it from $D$ and join it with all the
segments currently stored in $D$. If we use a balanced search tree for $D$
(e.g. a red-black tree), then inserting and removing an endpoint will cost us
$\BigO(\log n)$. As we have $2n$ endpoints, we arrive at a total of
$\BigO(2n \log 2n) = \BigO(n \log n)$. Generating all the output will take
$\BigO(k)$. If we use a hash table, insertion and removal of endpoints can be
done in $\BigO(1)$, for a total of $\BigO(n)$. As long as we make sure that
the entries in the hash table are linked or packed compactly (as in our
gapless hash map), this will have an overall complexity of $\BigO(k)$.


\paragraph*{\REV{Generalization:\,}}

Joins with predicates involving Allen or ISEQL relations are not exactly the
same as the one-dimensional line segment intersection. Nevertheless, the joins
can be mapped onto orthogonal line segment intersection, which is a special
case of two-dimensional line segment intersection that can also be done in
$\BigO(n \log n + k)$, \REV{with $k = n^2$ in the worst case}, using a
plane-sweeping algorithm that traverses the segments sorted by one
dimension~\cite{Cormen09}. This also explains why there were no further
developments for interval joins recently, as the state-of-the-art algorithms
achieve this complexity.  However, when generating the output, we cannot just
join a segment with all active ones, we need to check additional constraints:
two segments can overlap on the x-axis, but may or may not do so on the
y-axis. As we will see shortly, this has implications for the data structure
$D$.


\paragraph*{\REV{Complexity of Different Join Predicates:\,}}

Let us now have a closer look at the different join predicates. For all of
them, we need the relations $\Rel r$ and $\Rel s$ to be sorted. Either we keep
them in a Timeline Index or operate in a streaming environment, in which they
are already sorted, or we need to sort them in $\BigO(n \log n)$. The
non-parameterized and parameterized versions of \JoinName{start preceding},
\JoinName{end following}, and \JoinName{before} (which includes
\JoinName{meets} in its parameterized version) are not hard to analyze. They
all have a complexity of $\BigO(n \log n + k)$. For \JoinName{start
  preceding}, we maintain the active tuple set of $\Rel r$ in a gapless hash
map, which means $\BigO(1)$ for the insertion and removal of a single tuple,
or $\BigO(2n) = \BigO(n)$ in total. Additionally, whenever we encounter a
starting event of $\Rel s$, we generate result tuples, resulting in a total of
$\BigO(k)$ for generating all the output. For the parameterized version, we
merely shift the endpoints of the tuples in $\Rel r$. \JoinName{end following}
is very similar, the only differences being that we generate output when
encountering ending events of $\Rel s$ and for the parameterized variant, we
shift the starting points of $\Rel r$. \JoinName{before} is not much
different, we shift both events of tuples in $\Rel r$ and whenever we
encounter starting events of $\Rel s$, we generate the output.

We now turn to \JoinName{overlap} and \JoinName{during} joins, which we
implement using \JoinName{start preceding} (or \JoinName{end following})
joins; the same reasoning also holds for our implementation of the
\JoinName{equals}, \JoinName{starts}, and \JoinName{finishes} joins.
\REV{Processing a \JoinName{left overlap} or a \JoinName{reverse during} join},
we cannot just output the results in a straightforward way  
when encountering a starting
event in $\Rel s$ as before, as \REV{at this point we cannot determine whether
  two intervals are in a left-overlap or reverse-during relationship: the
  relationship between the starting events both look the same, we need to see
  the ending events to make a final decision}.
A similar argument holds for
implementing \JoinName{overlap} and \JoinName{during} joins with \JoinName{end
  following} joins: the role of the starting and ending events are switched in
this case. The textbook solution is to keep the intervals sorted by ending
events, e.g. in a tree. We can then search quickly for the qualifying tuples
in this tree and generate the output, resulting in an overall complexity of
$\BigO(n \log n + k)$.\footnote{Assuming that insertion and removal costs us
  $\BigO(\log n)$.} However, it is more difficult to do this in a
cache-friendly manner, as a tree traversal entails more random I/O than a
sequential scan.
\REV{Using a gapless hash map instead, we go through {\em all} the tuples in
  the active tuple set. Compared to the tree data structure, the processing of
  the join generates a larger intermediate result, as we join all intervals
  that satisfy an \JoinName{overlap} or \JoinName{during} join predicate. We
  filter out the tuples satisfying the predicate we are not interested in
  afterwards with a selection operator. Consequently, our approach has an
  overall complexity of $\BigO(n \log n + k')$ for \JoinName{overlap} and
  \JoinName{during} joins, with $k' geq k$. However, we utilize a sequential
  scan during the processing and as we will
see in the experimental evaluation, introducing random I/O into the traversal
of the active tuple set (like in a tree data structure)
starves the CPU and slows down the whole process by
two orders of magnitude.} On paper, our approach looks worse, but in practice
it outperforms the allegedly better method. 

\section{Experimental Evaluation}
\label{sec:experimental-evaluation}

\subsection{Setup}
\label{sec:setup}


\paragraph*{Environment:\,}

All algorithms were implemented in-memory in C++ by
the same author and compiled with GCC~4.9.4 
to 64-bit binaries using
the \texttt{\rule{0pt}{1ex}-O3} optimization flag.
We executed the code on a machine with two Intel
Xeon E5-2667 v3 processors under Linux. All experiments used 12-byte tuples
containing two 32-bit timestamp attributes ($T_s$ and $T_e$) and a 32-bit
integer payload. All experiments were repeated (also with bigger tuple sizes) on a seven-year-old Intel
Xeon X5550 processor and on a notebook processor i5-4258U, showing a similar
behavior.


\paragraph*{Algorithms:\,}

We compare our approach with the Leung-Muntz family of sweeping algorithms
\cite{leung_query_1989,leung_query_1990} and with an algorithm for generic
inequality joins, IEJoin \cite{khayyat_fast_2017}. We implemented the
Leung-Muntz algorithms in the most effective way, i.e., performing all stages
of the algorithm simultaneously, as recommended by the authors.  For a fair
comparison, we stored the set of started tuples in a Gapless List, adapting
the Gapless Hash Map technique (Section~\ref{sec:map}) to the Leung-Muntz
algorithms to boost their performance.
We implemented IEJoin using all optimizations from the original paper. Our
algorithms were implemented as described before, i.e., using abstractions and
lambda-functions.

The workload for all algorithms consisted
of accumulating the sum of $T_s$ attributes of the joined tuples.  For 
benchmarking, we implemented the tuples as structures and the relations as
\texttt{std::vector} containers. The Endpoint Index was implemented
analogously, using structures for the endpoints and a vector for the index.



\paragraph*{Synthetic Datasets:\,}
\label{sec:synthetic-datasets}

To show particular performance aspects of the algorithms we create synthetic
datasets with uniformly distributed starting points of the intervals in the
range of $[1,10^6]$.  The duration of the intervals is distributed
exponentially with rate parameter $\lambda$ (with an average duration
$1/\lambda$).  To perform a join, both relations in an individual workload
follow the same distribution, but are generated independently with a different
seed. In the experiments, for a specific value of
  $\lambda$, we varied the cardinality of the
generated relations.

\paragraph*{Real-World Datasets:\,}
\label{sec:rw-datasets}

We use five real-world datasets that differ in size and data distribution. The
main properties of them are summarized in Table~\ref{table:rw-datasets}. Here
$n$ is the number of tuples, $|r.T|$ is the tuple interval length, ``$r.T_s$ and $r.T_e$ domain''
is the size of the time domain of the dataset and ``$r.T_s$ and $r.T_e$ \#distinct'' is the number of
distinct time points in the dataset.

\begin{table}
\caption{Real-world dataset statistics}
\label{table:rw-datasets}
\centering
\AfterTableCaptionSpace
\begin{tabular}{@{}lrrrrrr@{}}
\toprule
    & &
    \multicolumn{3}{c}{$|r.T|$}  &
    \multicolumn{2}{c@{}}{$r.T_s$ and $r.T_e$} \\
\cmidrule(lr){3-5} \cmidrule(l){6-7}
dataset\!\!\!\!\! &    $n$ & min &    avg &    max &   domain & \#distinct \\
\midrule
flight  &  58\,k &  61 &   8\,k &  86\,k &   812\,k &      10\,k \\
inc     &  84\,k &   2 &    184 &    574 &     9\,k &     2.7\,k \\
web     & 1.2\,M &   1 &  60\,M & 352\,M &   352\,M &     110\,k \\
feed    & 3.7\,M &   1 &    432 & 8.5\,k &   8.6\,k &     5.6\,k \\
basf    & 5.3\,M &   1 & 127\,k &  16\,M &    16\,M &     760\,k \\
\bottomrule
\end{tabular}
\end{table}

The \emph{flight}
dataset~\cite{behrend_flight_data_2014} is a collection of international
flights for November 2014, start and end of the intervals represent plane
departure and arrival times with minute precision. 
The Incumbent (\emph{inc}) dataset~\cite{GendranoSSY98} records
the history of employees assigned to projects over a sixteen year period at a
granularity of days.
The \emph{web} dataset~\cite{webkit} records the
history of files in the SVN repository of the Webkit project over an eleven
year period at a granularity of seconds. The valid times indicate the periods
in which a file did not change. 
The \emph{feed} dataset records the history of measured
nutritive values of animal feeds over a 24 year period at a granularity of days; a
measurement remains valid until a new measurement for the same nutritive value
and feed becomes available \cite{dignos_overlap_2014}. 
Finally, rather than using time as a domain, the dataset \emph{basf} contains NMR
spectroscopy data describing the resonating frequencies of different atomic
nuclei \cite{Hel07}. As these frequencies can shift, depending on the bonds an
atom forms, they are defined as intervals.
For the experiments we used self-joins of these datasets, the only exception are the ``wi'' and ``fi'' workloads, where we joined the ``web'' and ``feed'' datasets with ``incumbent''.

\subsection{Experiments and Results}

\subsubsection{Cache Efficiency}


First, we look at the impact of improving the cache efficiency of the data
structure used for maintaining the active tuple set.  We investigate the
average latency of a \texttt{getnext} operation, which is crucial for
generating the result tuples.  We compare a linked hash map
(Section~\ref{sec:active-tuple-maps}), a gapless hash map
(Section~\ref{sec:map}), and a tree structure (mentioned in
Section~\ref{sec:theoretical}). The tree was implemented using a red-black
tree (std::map) from the C++ Standard Library.

We filled the data structures with various numbers of 32-byte tuples, then
randomly added and removed tuples to simulate the management of an active
tuple set. Afterwards, we performed several scans of the data structures.
Figure~\ref{fig:linked-vs-gapless-getnext}, shows the average latency of a
\texttt{getnext} operation depending on the number of tuples (note the
double-logarithmic scale).

\begin{figure}[htb]
\centering
\begin{tikzpicture}
\begin{loglogaxis}
[
        list performance plot,
        ylabel = {\texttt{getnext} latency, ns},
]
\addplot table[x=size, y=t]\dataPLists;
\addlegendentry{Tree}
\addplot table[x=size, y=l]\dataPLists;
\addlegendentry{Linked hash map}
\addplot table[x=size, y=u]\dataPLists;
\addlegendentry{Gapless hash map}
\end{loglogaxis}
\end{tikzpicture}
\vspace*{-.15cm}
\caption{Latency of \texttt{getnext} operation}
\label{fig:linked-vs-gapless-getnext}
\end{figure}
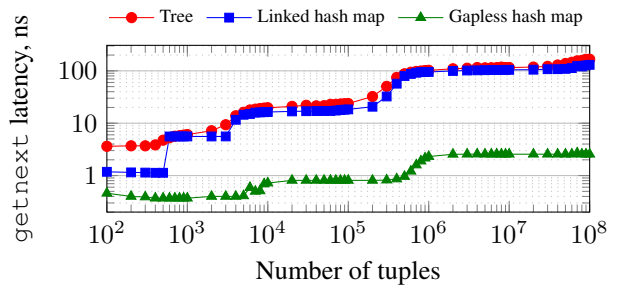

We see that the latency of a \texttt{getnext} operation is not constant but
grows depending on the memory footprint of the tuples. In order to find the
cause of this, we used the Performance Application Programming Interface
(PAPI) library to read out the CPU performance counters \cite{MoRa11}. When
looking at the average number of stalled CPU cycles (PAPI-RES-STL) per {\tt
  getnext} operation, we get a very similar
picture (see Figure~\ref{fig:linked-vs-gapless-getnext-res-stl}). Therefore,
the latency is clearly caused by the CPU memory subsystem.

\begin{figure}[htb]
\centering
\begin{tikzpicture}
\begin{loglogaxis}
[
        list performance plot,
        ylabel = {Stalled cycles},
        ymin = 0.3, ymax = 1000,
]
\addplot table[x=size, y=t-PAPI-RES-STL]\dataPLists;
\addlegendentry{Tree}
\addplot table[x=size, y=l-PAPI-RES-STL] \dataPLists;
\addlegendentry{Linked hash map}
\addplot table[x=size, y=u-PAPI-RES-STL] \dataPLists;
\addlegendentry{Gapless hash map}
\end{loglogaxis}
\end{tikzpicture}
\vspace*{-.15cm}
\caption{Stalled CPU cycles  per \texttt{getnext} operation}
\label{fig:linked-vs-gapless-getnext-res-stl}
\end{figure}
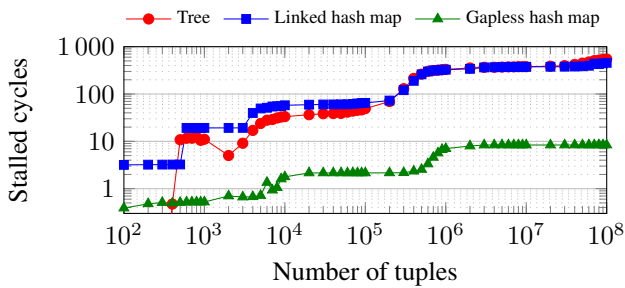

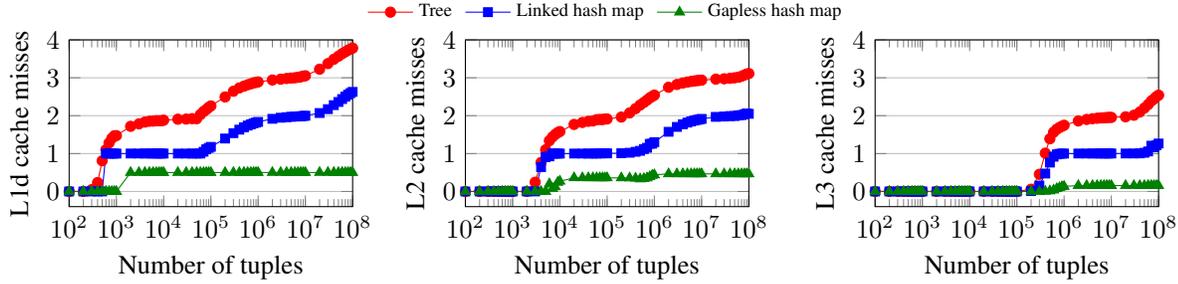
\begin{figure*}[htb]

\begin{tabular}{ccc}
\begin{tikzpicture}
\begin{loglogaxis}
[
        list performance plot small,
        ymode=normal,
        yticklabel=,
        ylabel = {L1d cache misses},
        ymax = 4,
]
\addplot table[x=size, y=t-PAPI-L1-DCM] \dataPLists;
\addplot table[x=size, y=l-PAPI-L1-DCM] \dataPLists;
\addplot table[x=size, y=u-PAPI-L1-DCM] \dataPLists;
\end{loglogaxis}
\end{tikzpicture}
&
\hspace*{-2em}\begin{tikzpicture}
\begin{loglogaxis}
[
        list performance plot small,
        ymode=normal,
        yticklabel=,
        ylabel = {L2 cache misses},
        ymax = 4,
]
\addplot table[x=size, y=t-PAPI-L2-TCM] \dataPLists;
\addlegendentry{Tree}
\addplot table[x=size, y=l-PAPI-L2-TCM] \dataPLists;
\addlegendentry{Linked hash map}
\addplot table[x=size, y=u-PAPI-L2-TCM] \dataPLists;
\addlegendentry{Gapless hash map}
\end{loglogaxis}
\end{tikzpicture}\hspace*{-3em}
&
\begin{tikzpicture}
\begin{loglogaxis}
[
        list performance plot small,
        ymode=normal,
        yticklabel=,
        ylabel = {L3 cache misses},
        ymax = 4,
]
\addplot table[x=size, y=t-PAPI-L3-TCM]\dataPLists;
\addplot table[x=size, y=l-PAPI-L3-TCM]\dataPLists;
\addplot table[x=size, y=u-PAPI-L3-TCM]\dataPLists;
\end{loglogaxis}
\end{tikzpicture}
\\
\end{tabular}
\vspace*{-.15cm}
\caption{Cache misses per \texttt{getnext} operation}
\label{fig:cache-miss}
\end{figure*}

In Figure~\ref{fig:linked-vs-gapless-getnext} we can easily identify
three distinct transitions. In case of a small number of tuples, all of them
fit into the L1d CPU cache (32~KB \REV{per core}) and we have a low latency. For the tree and
linked hash map, as the tuple count grows
towards \REV{500 tuples}, we start using the L2 cache (256~KB \REV{per core}) with a greater latency.
When we increase the number of tuples further and start reaching \REV{4000 tuples},
the data is mostly held in the L3 cache (20~MB \REV{in total, shared by all
  cores}) and, finally, after arriving at a tuple
count of around \REV{300\,000}, the tuples are mostly located in
RAM.\footnote{\REV{All CPUs have 32~KB and 256~KB per core for the L1d and L2 cache,
  respectively. The L3 cache for the Xeon X5550 is 8~MB and for the i5-4258u
  3~MB, which means that they reach the last phase earlier.}}
  We make a couple
of important observations. First, due to the more compact storage scheme of the
gapless hash map, the transitions set in later \REV{(at 5000, 10\,000, and
  600\,000 tuples, respectively)}. Second, the improvement gains
of the gapless hash map are considerable and can be measured in orders of
magnitude (note the logarithmic scale). Third, the latency of a
\texttt{getnext} operation for the gapless hash map plateaus at around 2.7~ns,
while the latency for the linked hash map and the tree reaches 100~ns.

Cache misses alone do not explain all of the
latency. Figure~\ref{fig:cache-miss} shows the average number of cache misses
for the L1d (PAPI-L1-DCM), the L2 (PAPI-L2-TCM), and the L3 cache
(PAPI-L3-TCM). While in general the average number of cache misses per {\tt
  getnext} operation is lower for the gapless hash map, the factor between the
data structures in terms of stalled CPU cycles is disproportionately higher
(please note the double-logarithmic scale in Figure~\ref{fig:linked-vs-gapless-getnext-res-stl}).
Also, the cache misses do not explain the left-most part of
Figure~\ref{fig:linked-vs-gapless-getnext-res-stl}, in which there are no
cache misses at all. The additional performance boost stems from out-of-order
execution.  Examining the
different (slightly simplified) versions of the machine code generated for
{\tt getnext} makes this clear. For the gapless hash map, the code looks like
this:

\begin{scriptsize}
\begin{verbatim}
loop:
  add   rax, [rdx]
  add   rdx, 32         ; pointer += 32 (increment)
  cmp   rcx, rdx
  jne   loop
\end{verbatim}
\end{scriptsize}

\noindent
while for the linked hash map we have the following picture
(we omit the code for the tree, as it is much more complex):

\begin{scriptsize}
\begin{verbatim}
loop:
  add   rax, [rdx]
  mov   rdx, [rdx + 32] ; pointer = pointer->field (dereference)
  cmp   rcx, rdx
  jne   loop
\end{verbatim}
\end{scriptsize}

\noindent
When scanning through a gapless hash map, we add a constant to the pointer,
which means that there is no data dependency between loop
iterations. Consequently, the CPU is able to predict the instructions that
will be executed in the future and can already start preparing them
out-of-order (i.e., issue cache misses up front for the referenced data) while
some of the instructions are still waiting for data from the L1 cache. For the
linked hash map and the tree the CPU has to wait until a pointer to the next
item has been dereferenced.  In summary, multiple parallel cache misses in a
sequential access pattern are processed much faster than isolated requests to
random memory locations.

We made another observation: there were no L1 instruction (L1i) cache
misses. The increase of L1d cache misses for the linked hash map and the tree
for large numbers of tuples is caused by TLB cache misses.

We obtained very similar results for different CPUs on
different machines (the diagrams shown here are for an Intel Xeon E5-2667 v3
processor), which led us to the conclusion that the techniques we employ will
generally improve the performance on CPU architectures with a cache hierarchy,
prefetching, and out-of-order execution. For the remainder of the experiments
we only consider the gapless hash map, as it clearly outperforms the linked
hash map.

\subsubsection{Lazy Joining}

For every tuple in $\mathbf s$, the basic JoinByS algorithm
(Section~\ref{sec:corealg}) scans the current set of active tuples in $\mathbf
r$. Using the improved LazyJoinByS algorithm from
Section~\ref{sec:lazyjoining}, we can reduce the number of scans
considerably. As long as we only encounter starting events of tuples
in $\mathbf s$ and no events caused by tuples in $\mathbf r$, we can delay the
scanning of the active tuple set of $\mathbf r$.

\paragraph*{Analyzing the Data:$\;$}

We now take a closer look at how frequently such uninterrupted sequences of
events of one relation appear. Figure~\ref{fig:srf} shows this data for the
table ``Incumbent'' from the real-world datasets when joining it with
itself. On the x-axis we have the length of uninterrupted sequences of
starting events and on the y-axis their relative frequency of appearance. In
60\% of the cases we have sequences of length ten or more, meaning that our
lazy joining technique can avoid a considerable number of scans on active tuple sets.

We found that starting events of intervals are generally not uniformly distributed in
real-world datasets, but tend to cluster around certain time points. This can
be recognized by looking at the number of distinct points in
Table~\ref{table:rw-datasets}. For example, for the ``Incumbent'' dataset, employees are
usually not assigned to new projects on random days, the assignments tend to
happen at the beginning of a week or month. For the ``Feed'' dataset,
multiple measurements (which are valid until the next one is made) are taken
in the course of a day, resulting in a whole batch of intervals starting at
the same time. The clustering is not just due to the relatively coarse
granularity (one day) of these two datasets. The ``Webkit'' repository dataset, which looks at intervals in which files are not modified, has a
granularity measured in milliseconds. Still we observe a clustering of
starting events: a commit usually affects and modifies several files. The
``Flight'' dataset, which has a granularity of minutes, also exhibits a
similar pattern in the form of batched departure times. Even for the frequency
data of the ``BASF'' dataset, the values for the start and end points of the
intervals seem to be clustered around multiples of one hundred.

\begin{figure}[htb]
\centering
\begin{tikzpicture}
\begin{axis}
[
  height=4cm,
  width=8cm,
  ymin=0,
  ymax=100,
  symbolic x coords={1,2,3,4,5,6,7,8,9,10+},
  xtick={1,2,3,4,5,6,7,8,9,10+},
  ybar,
  xlabel={Uninterrupted endpoint sequence length},
  ylabel={Relative frequency},
  yticklabel=\pgfmathprintnumber{\tick}\,\%,
  nodes near coords={\pgfmathprintnumber{\pgfplotspointmeta}\,\%},
  nodes near coords align={vertical},
  every node near coord/.append style={rotate=90, anchor=west, font=\scriptsize}
]
\addplot[fill=blue,draw=black] coordinates {(1,11) (2,6.4) (3,5.3) (4,4.4) (5,3.8) (6,2.6) (7,2.4) (8,2.3) (9,1.6) (10+,60)};
\end{axis}
\end{tikzpicture}
\vspace*{-.15cm}
\caption{Distribution of uninterrupted sequence lengths for self-join of the ``inc'' dataset}
\label{fig:srf}
\end{figure}
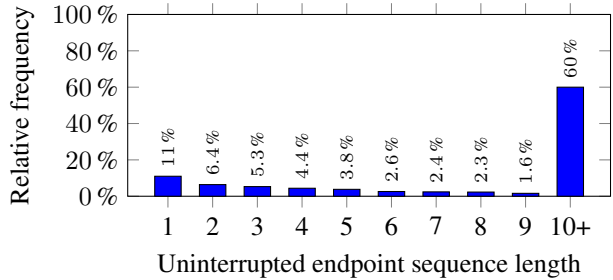

\paragraph*{Reduction Factor:$\;$}

The real performance implication is that LazyJoinByS executes fewer {\tt
  getnext} operations than JoinByS in such a scenario. The actual reduction depends not only on
the clusteredness of the events, but also on
the size of the corresponding active tuple set
and the buffer capacity reserved in LazyJoinByS. We define a \emph{{\tt
    getnext} operation reduction factor} ($\mathit{GNORF}$), changing
the cost for scanning through active tuple sets for the LazyJoinByS to
$\sfrac{k \cdot c_{getnext}}{\mathit{GNORF}}$, 
where $k$ is the cardinality of the result set.

For the self-join of the ``Incumbent'' dataset and for buffer capacity of 32
the $\mathit{GNORF}$ is equal to 23.6, which corresponds to huge savings in
terms of run time. We also calculated this statistic for self-joins of other
real-world datasets (``feed'': 31.2,  ``web'': 9.73, ``flight'': 7.14, ``basf'': 11.2).
Even for self-joins we get a considerable reduction factor: when encountering
multiple starting events with the same timestamp, we first deal with all those
of one relation before those of the other.

\paragraph*{Join Performance:$\;$}

%
%

Next we investigate the relative
performance of an actual join operation, employing JoinByS and LazyJoinByS for
an overlap join.
Figure~\ref{fig:basic-vs-cache-optimized-rw} depicts the results for
the ``Incumbent'' (inc), ``Webkit'' (web), ``Feed'' (feed), ``flight'', and ``basf'' datasets,
showing that LazyJoinByS outperforms JoinByS by up to a
factor of eight. Therefore, we only consider LazyJoinByS from here on.

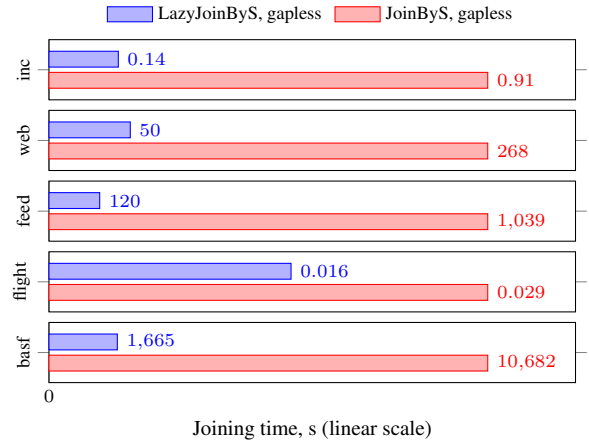
\begin{figure}[htb]
\centering
\begin{tikzpicture}
\begin{axis}
[
        real-world join bar plot small = {inc},
        xlabel = {},
        xticklabel=\empty,
        name=first,
        legend columns=-1
]
\pgfplotstablegetelem{0}{2u}\of\dataRW
\AddPlotCoord{inc}; \addlegendentry{JoinByS, gapless}
\pgfplotstablegetelem{0}{4u}\of\dataRW
\AddPlotCoord{inc}; \addlegendentry{LazyJoinByS, gapless}
\end{axis}
\begin{axis}
[
        real-world join bar plot small = {web},
        xlabel = {},
        xticklabel=\empty,
        /pgf/number format/precision=0,
        anchor=north west, at=(first.below south west),
        name=second,
]
\pgfplotstablegetelem{2}{2u}\of\dataRW
\AddPlotCoord{web};
\pgfplotstablegetelem{2}{4u}\of\dataRW
\AddPlotCoord{web};
\end{axis}
\begin{axis}
[
        real-world join bar plot small = {feed},
        xlabel = {},
        xticklabel=\empty,
        /pgf/number format/precision=0,
        anchor=north west, at=(second.below south west),
        name=third,
]
\pgfplotstablegetelem{3}{2u}\of\dataRW
\AddPlotCoord{feed};
\pgfplotstablegetelem{3}{4u}\of\dataRW
\AddPlotCoord{feed};
\end{axis}
\begin{axis}
[
        real-world join bar plot small = {flight},
        xlabel = {},
        xticklabel=\empty,
        /pgf/number format/precision=3,
        anchor=north west, at=(third.below south west),
        name=fourth,
]
\pgfplotstablegetelem{1}{2u}\of\dataRW
\AddPlotCoord{flight};
\pgfplotstablegetelem{1}{4u}\of\dataRW
\AddPlotCoord{flight};
\end{axis}
\begin{axis}
[
        real-world join bar plot small = {basf},
        /pgf/number format/precision=0,
        anchor=north west, at=(fourth.below south west),
]
\pgfplotstablegetelem{4}{2u}\of\dataRW
\AddPlotCoord{basf};
\pgfplotstablegetelem{4}{4u}\of\dataRW
\AddPlotCoord{basf};
\end{axis}
\end{tikzpicture}
\vspace*{-.15cm}
\caption{JoinByS vs LazyJoinByS, real-world data}
\label{fig:basic-vs-cache-optimized-rw}
\end{figure}

\newcommand*{\AddSpeedupPlot}[1] 
{
    \ReadTable{#1}
    \AddPlot[]{x = dataset, y expr = (\thisrow{lm} / \thisrow{danila}) }
}

\newcommand*{\AddSpeedupPlotTreeRed}[1] 
{
    \ReadTable{#1}
    \AddPlot[dashed,red,mark=*,mark size = 1.92]{x = dataset, y expr = (\thisrow{tree} / \thisrow{danila}) }
}

\newcommand*{\AddSpeedupPlotTreeBlue}[1] 
{
    \ReadTable{#1}
    \AddPlot[dashed,blue,mark=square*,mark size = 1.66]{x = dataset, y expr = (\thisrow{tree} / \thisrow{danila}) }
}

\newcommand*{\AddSpeedupPlotTreeGreen}[1] 
{
    \ReadTable{#1}
    \AddPlot[dashed,MyGreen,mark=triangle*,mark size = 2.13]{x = dataset, y expr = (\thisrow{tree} / \thisrow{danila}) }
}

\newcommand*{\AddSpeedupPlotTree}[1] 
{
    \ReadTable{#1}
    \AddPlot[dashed]{x = dataset, y expr = (\thisrow{tree} / \thisrow{danila}) }
}

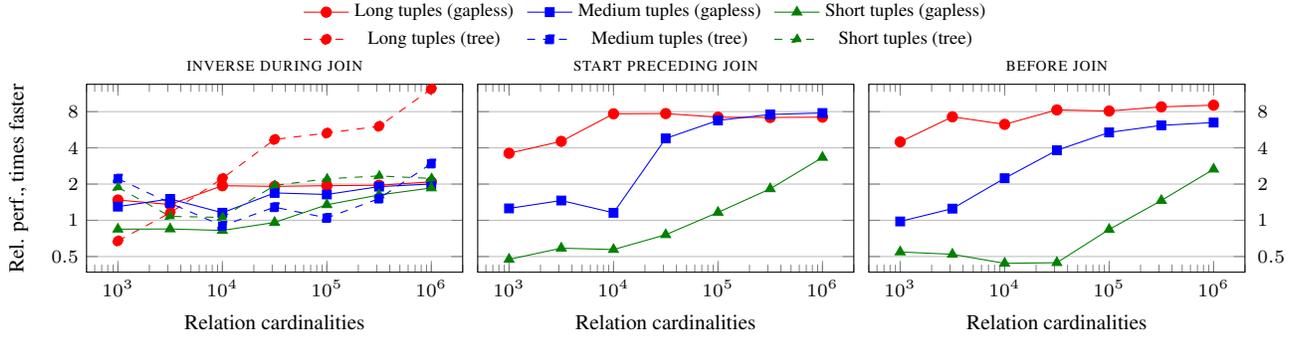
\begin{figure*}
  \centering
\ref{legend-exp}\\
\begin{tikzpicture}
\begin{groupplot}
[
    three plots,
    xmode = log,
	xlabel = {Relation cardinalities},
	ylabel = {Rel. perf., times faster},
    ymin = 0.5, ymax = 10,
    xmin = 1e3, xmax = 1e6,
    legend columns = 3,
    enlarge x limits = 0.1,
    enlarge y limits = 0.1,
    ymode = log, log basis y = 2, log y ticks with fixed point base 2,
]
\nextgroupplot[
    legend to name = legend-exp, 
    title = {\JoinName{Inverse During Join}}
]
\AddSpeedupPlot{exp-reverse-during-w1e6.txt}
\addlegendentry{Long tuples (gapless)}
\AddSpeedupPlot{exp-reverse-during-w1e4.txt}
\addlegendentry{Medium tuples (gapless)}
\AddSpeedupPlot{exp-reverse-during-w1e2.txt}
\addlegendentry{Short tuples (gapless)}
\AddSpeedupPlotTreeRed{exp-reverse-during-w1e6.txt}
\addlegendentry{Long tuples (tree)}
\AddSpeedupPlotTreeBlue{exp-reverse-during-w1e4.txt}
\addlegendentry{Medium tuples (tree)}
\AddSpeedupPlotTreeGreen{exp-reverse-during-w1e2.txt}
\addlegendentry{Short tuples (tree)}
\nextgroupplot[
    multiple plots middle,
    title = {\JoinName{Start Preceding Join}}
]
\AddSpeedupPlot{exp-start-preceding-w1e6.txt}
\AddSpeedupPlot{exp-start-preceding-w1e4.txt}
\AddSpeedupPlot{exp-start-preceding-w1e2.txt}
\nextgroupplot[   
    multiple plots right,
    title = {\JoinName{Before Join}}
]
\AddSpeedupPlot{exp-before-join-w1e6.txt}
\AddSpeedupPlot{exp-before-join-w1e4.txt}
\AddSpeedupPlot{exp-before-join-w1e2.txt}
\end{groupplot}
\end{tikzpicture}
\caption{Performance of our solution w.r.t.\ Leung-Muntz algorithm, synthetic data}
\label{fig:exp}
\end{figure*}

\renewcommand*{\AddSpeedupPlot}[1] 
{
    \ReadTable{#1}
    \AddPlot[]{x = dataset, y expr = (\thisrow{ie} / \thisrow{danila}) }
}

\begin{figure*}[htb]
  \centering
  \ref{legend-ie-exp}
\begin{tikzpicture}
\begin{groupplot}
[
    three plots,
    xmode = log,
	xlabel = {Relation cardinalities},
	ylabel = {Rel. perf., times faster},
    ymin = 1, ymax = 1e5,
    xmin = 1e3, xmax = 1e6,
    legend columns = 3,
    enlarge x limits = 0.1,
    ymode = log, 
]
\nextgroupplot[
    legend to name = legend-ie-exp, 
    title = {\JoinName{Inverse During Join}}
]
\AddSpeedupPlot{ie-exp-reverse-during-w1e6.txt}
\addlegendentry{Long tuples}
\AddSpeedupPlot{ie-exp-reverse-during-w1e4.txt}
\addlegendentry{Medium tuples}
\AddSpeedupPlot{ie-exp-reverse-during-w1e2.txt}
\addlegendentry{Short tuples}
\nextgroupplot[
    multiple plots middle,
    title = {\JoinName{Start Preceding Join}}
]
\AddSpeedupPlot{ie-exp-start-preceding-w1e6.txt}
\AddSpeedupPlot{ie-exp-start-preceding-w1e4.txt}
\AddSpeedupPlot{ie-exp-start-preceding-w1e2.txt}
\nextgroupplot[
    multiple plots right,
    title = {\JoinName{Before Join}}
]
\AddSpeedupPlot{ie-exp-before-join-w1e6.txt}
\AddSpeedupPlot{ie-exp-before-join-w1e4.txt}
\AddSpeedupPlot{ie-exp-before-join-w1e2.txt}
\end{groupplot}
\end{tikzpicture}
\caption{Performance of our solution w.r.t.\ IEJoin, synthetic data}
\label{fig:ie-exp}
\end{figure*}
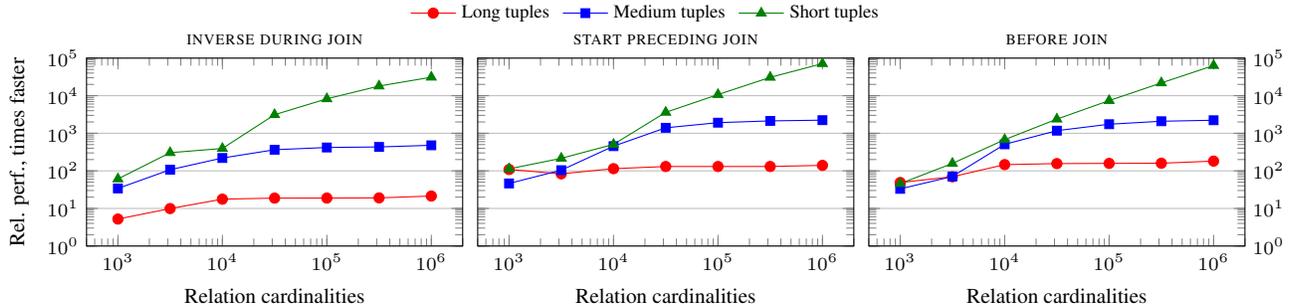

\subsubsection{Scalability}

To test the scalability of our algorithms we compared them to the Leung-Muntz
and IEJoin algorithms while varying the cardinality of the synthetic
datasets. 
For the IEJoin we used the algebraic expressions in Table~\ref{table:allen-relations}.
Due to space constraints, we limit ourselves to three characteristic joins
(join-only, join with selection, and parameterized join with map operators):
\JoinName{start preceding} (\Doodle{\StartPreceding}), \JoinName{inverse
  during} (\Doodle{\Reverse{\During}}), 
and \JoinName{before} join (\Doodle{\Before}), where
$\delta$ is set to the average tuple length in the outer relation.
We tested the algorithms using short, medium-length and long tuples with
average lengths of $0.5 \cdot 10^2$, $0.5 \cdot 10^4$, and $0.5 \cdot 10^6$
time points, respectively.
The speedup of our approach compared to Leung-Muntz
and IEJoin is shown in Figures~\ref{fig:exp} and~\ref{fig:ie-exp},
respectively. We see that our solution
quickly becomes faster than the Leung-Muntz algorithms and that the difference
grows with increasing numbers of tuples and their lengths. Only for
small relations and short tuples, Leung-Muntz is faster.
\REV{Leung-Muntz is a simpler algorithm, so for light workloads, i.e.,
  small relations and short intervals (resulting in smaller result sets),
  it shows a good performance as it does not have an initialization overhead.
  However, it does not scale as well as our algorithm.}
In the left-most
diagram of Figure~\ref{fig:exp} (\JoinName{inverse during join}), we also show the
difference between using gapless hash maps and trees for managing the active
tuple set. We only do so for the \JoinName{inverse during join}, as for the other
joins a tree would only add overhead without any benefits. For the
\JoinName{inverse during join}, with a tree we generate only valid tuples, meaning
that we do not need a selection operator as needed for the gapless hash
map. While the tree-based active tuple set seems to pay off for long tuples
(meaning larger active tuple sets), for shorter tuples the situation is not
that clear. Consequently, we propose to use gapless hash maps, as this avoids
the implementation and integration of an additional data structure that is
only useful for some interval relations and even then does not always show
superior performance. For the remainder of
Section~\ref{sec:experimental-evaluation}, we restrict ourselves to gapless
hash maps.

For the IEJoin, the performance differs by one to several orders of magnitude
depending on relation cardinalities and tuple lengths.
Even though the IEJoin is highly optimized, it still has quadratic complexity
and cannot compete with specialized algorithms. Therefore, we drop it from the
further investigation. Because the Leung-Muntz and the IEJoin algorithms do
not scale well, we stopped running experiments for larger relation
cardinalities when they took a few hours to conduct. We also
restrict ourselves to the \JoinName{inverse during} join from now on, since Leung-Muntz
showed the best performance for it.

\subsubsection{Real-World Workloads}

\begin{figure}[t]
\begin{tikzpicture}
\begin{axis}
[
    global plot style,
    real world bar plot,
	ylabel = {Rel. perf., times faster},
    ymin = 1, ymax = 128,
    ymode = log, log basis y = 2, log ticks with fixed point,
]
\ReadTable{rw-reverse-during.txt}
\AddPlot[]{x = dataset, y expr = (\thisrow{lm} / \thisrow{danila}) }
\addlegendentry{\JoinName{during}}
\ReadTable{rw-start-preceding.txt}
\AddPlot[]{x = dataset, y expr = (\thisrow{lm} / \thisrow{danila}) }
\addlegendentry{\JoinName{start preceding}}
\ReadTable{rw-before-join.txt}
\AddPlot[]{x = dataset, y expr = (\thisrow{lm} / \thisrow{danila}) }
\addlegendentry{\JoinName{before}}
\end{axis}
\end{tikzpicture}
\caption{Performance of our solution w.r.t.\ Leung-Muntz, real-world data}
\label{fig:rw}
\end{figure}
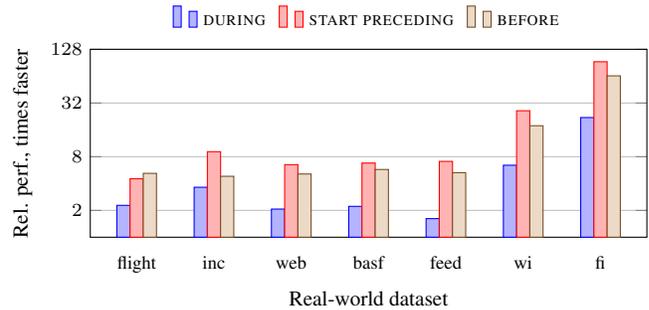

We repeated the experiments on the real-world workloads.  
The speedup is shown in Figure~\ref{fig:rw}.
Here the performance difference is two orders of magnitude in some cases. 
On the one hand, this is due to the lazy joining cache optimization, which is more effective for the real-world datasets (cf.~\cite{piatov_interval_2016}). 
On the other hand, the heuristics used in the Leung-Muntz algorithm work worse for real-worlds workloads and especially those where the relation cardinalities differ substantially. 


\REV{
\subsubsection{Comparison with bgFS}

Our approach and bgFS~\cite{bouros_forward_2017} follow different paradigms for
processing the data: backward scanning versus forward scanning. The iterator
framework we utilize has been geared completely towards the backward scanning
paradigm, allowing us to introduce changes to endpoint timestamps on the
fly. This makes it challenging to integrate bgFS into our iterator framework
effectively. Clearly, we can add shifted intervals to the tuples in the
relations before executing an bgFS join. However, this requires and additional
sweep over the relations, eating up the efficiency gained by forward scanning
the relations. On top of that, bgFS will start producing output tuples at the
very end of the processing time. Figure~\ref{fig:comparisonwithfs} shows the run time of
processing a (general) \JoinName{before} join using bgFS and our approach (this was
run on an i7-4870HQ CPU with four cores, 32~KB and 256~KB per core for the L1d
and L2 cache, respectively, and 6~MB L3 cache).
}

\begin{figure}[htb]
\centering
\begin{tikzpicture}
\begin{axis}
[
        real-world join bar plot small = {inc},
        xlabel = {},
        xticklabel=\empty,
        name=first,
        legend columns=-1
]
\pgfplotstablegetelem{0}{JoinByS}\of\dataFS
\AddPlotCoord{inc}; \addlegendentry{JoinByS}
\pgfplotstablegetelem{0}{bgFS}\of\dataFS
\AddPlotCoord{inc}; \addlegendentry{bgFS}
\end{axis}
\begin{axis}
[
        real-world join bar plot small = {web},
        xlabel = {},
        xticklabel=\empty,
        /pgf/number format/precision=0,
        anchor=north west, at=(first.below south west),
        name=second,
]
\pgfplotstablegetelem{3}{JoinByS}\of\dataFS
\AddPlotCoord{web};
\pgfplotstablegetelem{3}{bgFS}\of\dataFS
\AddPlotCoord{web};
\end{axis}
\begin{axis}
[
        real-world join bar plot small = {inc-web},
        xlabel = {},
        xticklabel=\empty,
        anchor=north west, at=(second.below south west),
        name=third,
]
\pgfplotstablegetelem{1}{JoinByS}\of\dataFS
\AddPlotCoord{inc-web};
\pgfplotstablegetelem{1}{bgFS}\of\dataFS
\AddPlotCoord{inc-web};
\end{axis}
\begin{axis}
[
        real-world join bar plot small = {web-inc},
        xticklabel=\empty,
        anchor=north west, at=(third.below south west),
        name=fourth,
]
\pgfplotstablegetelem{2}{JoinByS}\of\dataFS
\AddPlotCoord{web-inc};
\pgfplotstablegetelem{2}{bgFS}\of\dataFS
\AddPlotCoord{web-inc};
\end{axis}
\end{tikzpicture}
\vspace*{-.15cm}
\caption{Comparison of JoinByS with bgFS, real-world data}
\label{fig:comparisonwithfs}
\end{figure}
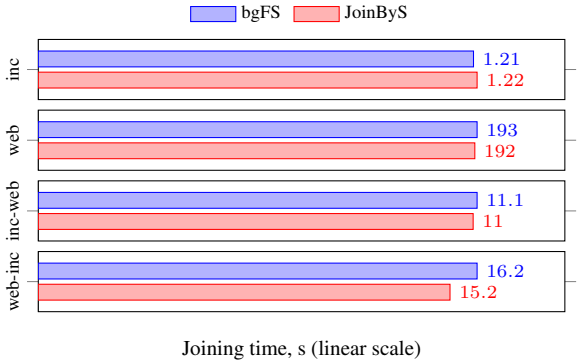

\subsubsection{Selection Efficiency}

To explore the source of the performance difference between the algorithms, we
tested the selectivity of the selection operation that is applied after the
join in the Leung-Muntz algorithms and, in some cases, in ours. 
The results for the \JoinName{inverse during join} are shown in Figure~\ref{fig:selectivities}. 
The other joins exhibit a similar behavior. 
We see that both algorithms are keeping the sizes of the working sets similar. 
Our algorithm is slightly more effective, but insufficiently so to explain the performance difference.
We look at the real cause in the next section.

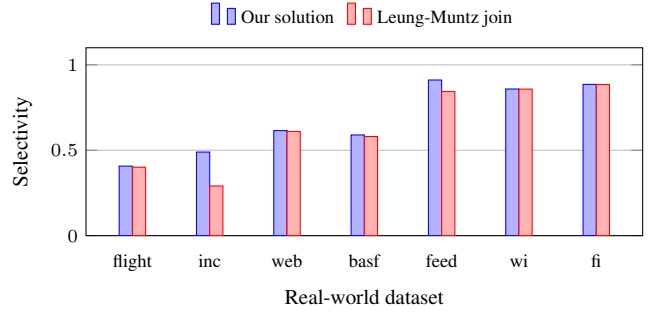
\begin{figure}[t]
\begin{tikzpicture}
\begin{axis}
[
    global plot style,
    real world bar plot,
	ylabel = {Selectivity},
    ymin = 0, ymax = 1.1,   
]
\ReadTable{counters-rw-reverse-during.txt}
\AddPlot[]{x = dataset, y = my-sel}
\addlegendentry{Our solution}
\AddPlot[]{x = dataset, y = lm-sel}
\addlegendentry{Leung-Muntz join}
\end{axis}
\end{tikzpicture}
\caption{Selectivity of the filtering selection operator after the main join, \JoinName{inverse during join}}
\label{fig:selectivities}
\end{figure}

\subsubsection{Comparison Count}

In this experiment we measured the number of tuple endpoint comparison
operations (e.g., ``$r.T_s < s.T_s$'').  The results for the \JoinName{inverse
  during join} are shown in Figure~\ref{fig:comparisons}.  We see that the
difference in the number of comparisons is huge.  The Leung-Muntz algorithm
performs many more comparisons because it has to heuristically estimate the
next tuple to be read and to perform the garbage collection of the outdated
tuples.  The selection operation of the Leung-Muntz algorithm requires two
comparisons.  Our algorithm, on the other hand, requires a single comparison
in the selection operation for the \JoinName{inverse during join}, and no
tuple comparisons at all for the \JoinName{before} and \JoinName{start
  preceding join}.

\begin{figure}
\begin{tikzpicture}
\begin{axis}
[
    global plot style,
    real world bar plot,
	ylabel = {Comparison count},
    ymode = log,
]
\ReadTable{counters-rw-reverse-during.txt}
\AddPlot[]{x = dataset, y = my-comp}
\addlegendentry{Our solution}
\AddPlot[]{x = dataset, y = lm-comp}
\addlegendentry{Leung-Muntz join}
\end{axis}
\end{tikzpicture}
\caption{Join comparison counts, real-world data}
\label{fig:comparisons}
\end{figure}
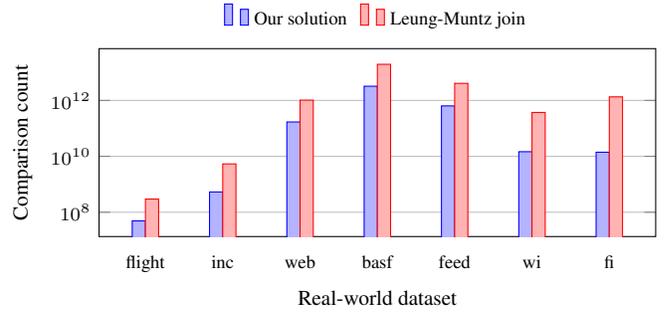

%
%
%

\subsubsection{Latency}

In this experiment we measure the delay in producing output tuples of the
Leung-Muntz algorithms. A low latency is an important feature
for event detection systems.
While our algorithms generates output as soon as
possible, when all of the required endpoints have been spotted, the
Leung-Muntz has a delay caused by the fact that it requires streams of
complete and ordered tuples as the input.  The average latency (expressed in
average tuple lengths, as the different data sets have vastly different
granularities) is shown in Figure~\ref{fig:latency}.  Depending on the
workload the differences in latency can in some cases reach ten or even a
hundred times the average tuple length.

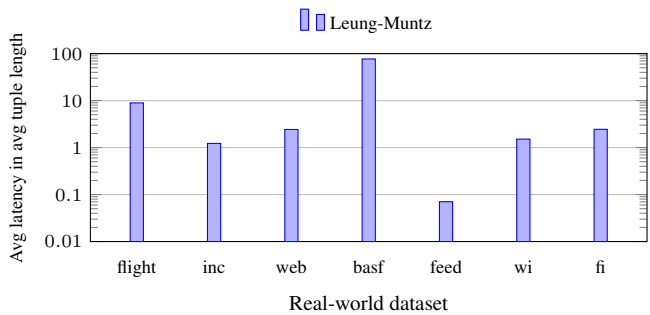
\begin{figure}
\begin{tikzpicture}
\begin{axis}
[
    global plot style,
    real world bar plot,
	ylabel = {\scriptsize{Avg latency in avg tuple length}},
    ymin = 0.01, ymax = 100,
    ymode = log, log origin = infty, log ticks with fixed point,
]
\ReadTable{latency.txt}
\AddPlot[]{x = dataset, y expr = \thisrow{latency-2-avg} / \thisrow{r-avg-len} }
\addlegendentry{Leung-Muntz}
\end{axis}
\end{tikzpicture}
\caption{Algorithm reporting latency, \RelationName{reverse during join}, real-world data}
\label{fig:latency}
\end{figure}

\section{Conclusions and Future Work}
\label{sec:conclusions}

We developed a family of effective, robust and cache-opti\-mized plane-sweeping
algorithms for interval joins on different interval relationship predicates
such as Allen's or parameterized ISEQL relations. The algorithms can be used in
temporal databases, exploiting the Timeline Index, which made its
way into a prototype of a commercial temporal RDBMS as the main universal
index supporting temporal joins, temporal aggregation and time travel. We thus
extend the set of operations supported by this index. Our solution is
based on a flexible framework, that allows combining its components in
different ways to elegantly and efficiently express any interval join in terms
of a single core function. Additionally, our approach makes good use of the
features of contemporary hardware, utilizing the cache infrastructure well.

We compared the performance of our solution with the
state-of-the-art in interval joins on Allen's predicates---the Leung-Muntz
and the IEJoin algorithm. The results show that our solution is several times faster,
scales better and is more stable. Another major advantage of our approach is that it can be directly
applied to real-time stream event processing, as it will report the results as
soon as logically possible for the applied predicate, without necessarily waiting for
intervals to finish. The Leung-Muntz algorithm
has to wait for the tuples to finish before processing them. Moreover, the requirement for tuples to be processed chronologically allows any unfinished tuple to block the processing of all following tuples.
The IEJoin is also not suitable for a streaming environment: it needs the
complete relations to work.

For future work, we want to explore the possibilities of embedding our
solution into a real-time complex event processing framework. In particular,
we want to combine the results of multiple joins to detect patterns within $n$
streams of events of different type. \REV{Additional research directions are
  working out the details of parallelizing our approach and the handling of
  multiple predicates in a join operation.}


\bibliographystyle{abbrv}      
\bibliography{vldbj-iseql-article}

\appendix

\iftechreport{
\section{Correctness of Rewrites}
\label{sec:rewrites}

Joining the tuples in the relations \Rel r and \Rel s such that they satisfy
the Allen and ISEQL relations shown in Table~\ref{table:allen-relations}
results in a tuple set 
$
RS_{name} = \SetBuilder{
        r \times s
    }{
        r \in \Rel r, s \in \Rel s: P(r,s)
    }
$
where $name$ is the name of the relation in the first column and $P(r,s)$ is
the formal definition in the third column. We follow the same order as in
Section~\ref{sec:formalization}.

\subsection{Non-parameterized Joins}

\paragraph*{ISEQL Start Preceding and End Following Joins:\,}

These follow directly from the definition of our interval-timestamp joins 
$\JoinByTssm^\theta$  and $\JoinByTesm^\theta$.

\paragraph*{Overlap and During Joins:\,}

For the \JoinName{left overlap join}, the tuples have to satisfy $P(r,s) = 
r.T_s \le s.T_s < r.T_e \le s.T_e$. The inequalities $r.T_s \le s.T_s < r.T_e$
are enforced by the join $\JoinByTssm^\le$, the inequality $r.T_e \le s.T_e$ by
the selection predicate. For the right overlap join, this works analogously
using the join $\JoinByTesm^\le$ and a selection. 

For the (reverse) \JoinName{during join}, only the inequality enforced by the
selection operator changes.

\paragraph{Before Join:\,}

By replacing $T_s$ with $T_e$ and $T_e$ with $\infty$ in every tuple in \Rel r
and then running a $\JoinByTssm^\le$ join, we get tuples that satisfy the
predicate $r.T_e \le s.T_s < \infty$, which is equivalent to the predicate for
the \JoinName{before} relation. For the Allen relation we use 
$\JoinByTssm^<$ instead.

\paragraph*{Meets Join:\,}

Applying the map operators and running a $\JoinByTssm^\le$ join, $P(r,s)$
becomes $r.T_e \le s.T_s < r.T_e + 1$. Since we use integer timestamps, this
means that $r.T_e = s.T_s$.

\paragraph*{Equals and Starts Joins:\,}

For the \JoinName{equals join} the $\JoinByTssm^\le$ join enforces the predicate
$r.T_s \le s.T_s < r.T_s + 1$, which becomes $r.T_s = s.T_s$ due to the
integer timestamps, while the selection does so for $r.T_e = s.T_e$. For
the \JoinName{starts join} the predicate enforced by the selection operator
changes accordingly.

\paragraph*{Finishes Join:\,}

For the \JoinName{finishes join} we use the $\JoinByTesm^\le$ operator arriving
at $r.T_e - 1 < s.T_e \le r.T_e$, which means that $r.T_e = s.T_e$. Together
with the selection predicate of $s.T_s < r.T_s$, we complete the predicate for
the \JoinName{finishes join}.

\subsection{Parameterized Joins}

\paragraph*{ISEQL Start Preceding and End Following Joins:\,}

The $\JoinByTssm^\le$ join (together with the map operators) guarantees us that 
$r.T_s \le s.T_s < \min(r.T_e, r.T_s + \delta + 1)$. Assume that the minimum
is $r.T_e$, which means that $r.T_e \le r.T_s + \delta + 1$. The first part of
$P(r,s)$, $r.T_s \le s.T_s < r.T_e$ follows directly. We also know that
$s.T_s < r.T_e \le  r.T_s + \delta + 1$ and due to integer timestamps can
conclude that $s.T_s - r.T_s \le \delta$. 

For the other case, the minimum is  $r.T_s + \delta + 1$, which means 
$r.T_s + \delta + 1 \le r.T_e$. Due to integer timestamps, it immediately
follows that $s.T_s \le r.T_s + \delta$. We also know that
$r.T_s \le s.T_s < r.T_s + \delta + 1 \le r.T_e$ and therefore 
$r.T_s \le s.T_s < r.T_e$. 

The proof for the \JoinName{end following join} follows along similar lines.

\paragraph*{Before Join:\,}

The  $\JoinByTssm^\le$ join enforces $r.T_e \le s.T_s < r.T_e + \delta + 1$,
which due to integer timestamps becomes
$r.T_e \le s.T_s \le r.T_e + \delta$ (which is equivalent to 
$r.T_e \le s.T_s \wedge s.T_s - r.T_e \le \delta$).

\paragraph*{Overlap and During Joins:\,}

Here we only show the proof for the \JoinName{left overlap join}, the proofs
for the remaining operators follow the same pattern.

With the $\JoinByTssm^\le$ join we already make sure that $r.T_s \le s.T_s$ and
with the selection operator that $r.T_e \le s.T_e$ and $s.T_e \le r.T_e
+ \varepsilon \Leftrightarrow s.T_e - r.T_e \le \varepsilon$. If $r.T_e \le
r.T_s + \delta + 1$ then $s.T_s < r.T_e$ follows from the $\JoinByTssm^\le$ as
well and because $s.T_s < r.T_e \le r.T_s + \delta$, we know that 
$s.T_s < r.T_s + \delta$, which is equivalent to $s.T_s - r.T_e \le \delta$
for integer intervals. If $r.T_s + \delta + 1 \le r.T_e$, then 
$s.T_s < r.T_s + \delta + 1 \Leftrightarrow s.T_s - r.T_s \le \delta$ (for
integer intervals) follows from the $\JoinByTssm^\le$ and because
$s.T_s < r.T_s + \delta + 1 \le r.T_e$, we also know that 
$s.T_s < r.T_e$. 

}

\section{Implementation of Iterators}
\label{appendix:iterators}

In this section we illustrate the inner workings of the different iterators
and show how they can be implemented. 
If we have an instance of an Endpoint Iterator, \texttt{iterator}, then to
traverse all endpoints it represents, we can use the following pseudocode:

\begin{footnotesize}
\begin{verbatim}
while not iterator.isFinished do
    output(iterator.getEndpoint)
    iterator.moveToNextEndpoint
end
\end{verbatim}
\end{footnotesize}
\vspace{-.5cm}

\subsection{Index Iterator}

We use the \texttt{std::vector} container of the C++ Standard
Template Library (STL) as the implementation of the Endpoint Index, resulting
in the following code:

\begin{itemize}
\footnotesize
\item \texttt{IndexIterator(endpointIndex)}:\\
    \hspace*{1em}\texttt{this.it  = endpointIndex.begin();}\\
    \hspace*{1em}\texttt{this.end = endpointIndex.end();}

\item \texttt{getEndpoint}:\\
    \hspace*{1em}\texttt{return *it;}

\item \texttt{moveToNextEndpoint}:\\
    \hspace*{1em}\texttt{++it;}

\item \texttt{isFinished}:\\
    \hspace*{1em}\texttt{return it == end;}
\end{itemize}
\vspace{-.5cm}

\subsection{Filtering Iterator}

\begin{itemize}
\footnotesize
\item \texttt{FilteringIterator(iterator, type)}:\\
    \hspace*{1em}\texttt{this.iterator = iterator;}\\
    \hspace*{1em}\texttt{this.type = type;}\\
    \hspace*{1em}\texttt{while getEndpoint.type $\ne$ type do}\\
    \hspace*{1em}\texttt{\hspace{2em}moveToNextEndpoint;}

\item \texttt{getEndpoint}:\\
    \hspace*{1em}\texttt{return iterator.getEndpoint;}

\item \texttt{moveToNextEndpoint}:\\
    \hspace*{1em}\texttt{do}\\
    \hspace*{1em}\texttt{\hspace{2em}iterator.moveToNextEndpoint;}\\
    \hspace*{1em}\texttt{while not isFinished and}\\
    \hspace*{1em}\texttt{\phantom{while }getEndpoint.type $\ne$ type;}

\item \texttt{isFinished}:\\
    \hspace*{1em}\texttt{return iterator.isFinished;}
\end{itemize}
\vspace{-.5cm}

\subsection{Shifting Iterator}

\begin{itemize}
\footnotesize
\item\texttt{ShiftingIterator(iterator, delta, type)}:\\
    \hspace*{1em}\texttt{this.iterator = iterator;}\\
    \hspace*{1em}\texttt{this.delta = delta;}\\
    \hspace*{1em}\texttt{this.type = type;}

\item \texttt{getEndpoint}:\\
    \hspace*{1em}\texttt{var endpoint = iterator.getEndpoint;}\\
    \hspace*{1em}\texttt{endpoint.timestamp += delta;}\\
    \hspace*{1em}\texttt{endpoint.type = type;}\\
    \hspace*{1em}\texttt{return endpoint;}

\item \texttt{moveToNextEndpoint}:\\
    \hspace*{1em}\texttt{iterator.moveToNextEndpoint;}

\item \texttt{isFinished}:\\
    \hspace*{1em}\texttt{return iterator.isFinished;}
\end{itemize}
\vspace{-.5cm}


\subsection{Merging Iterator}

\begin{itemize}
\footnotesize
\item \texttt{MergingIterator(iterator1, iterator2)}:\\
    \hspace*{1em}\texttt{this.it1 = iterator1;}\\
    \hspace*{1em}\texttt{this.it2 = iterator2;}\\
    \hspace*{1em}\texttt{moveToNextEndpoint;}

\item \texttt{getEndpoint}:\\
    \hspace*{1em}\texttt{return this.endpoint;}

\item \texttt{moveToNextEndpoint}:\\
    \hspace*{1em}\texttt{if it2.isFinished or not it1.isFinished}\\
    \hspace*{1em}\texttt{\phantom{if }and it1.getEndpoint < it2.getEndpoint}\\ \hspace*{1em}\texttt{then}\\
    \hspace*{1em}\texttt{\hspace{2em}this.endpoint = it1.getEndpoint;}\\
    \hspace*{1em}\texttt{\hspace{2em}it1.moveToNextEndpoint;}\\
    \hspace*{1em}\texttt{else}\\
    \hspace*{1em}\texttt{\hspace{2em}this.endpoint = it2.getEndpoint;}\\
    \hspace*{1em}\texttt{\hspace{2em}it2.moveToNextEndpoint;}\\
    \hspace*{1em}\texttt{end}

\item \texttt{isFinished}:\\
    \hspace*{1em}\texttt{return it1.isFinished and it2.isFinished;}
\end{itemize}
\vspace{-.5cm}

\REV{
\subsection{First End Iterator}
\label{sec:firstend}

\begin{itemize}
\footnotesize
\item \texttt{FirstEndIterator(iterator)}:\\
    \hspace*{1em}\texttt{this.iterator = iterator;}\\
    \hspace*{1em}\texttt{this.hs = new HashSet;}\\

\item \texttt{getEndpoint}:\\
    \hspace*{1em}\texttt{return this.endpoint;}
\item \texttt{moveToNextEndpoint}:\\
    \hspace*{1em}\texttt{do}\\
    \hspace*{1em}\texttt{\hspace{2em}iterator.moveToNextEndpoint;}\\
    \hspace*{1em}\texttt{\hspace{2em}if getEndpoint.type = end then}\\
    \hspace*{1em}\texttt{\hspace{4em}if getEndpoint.tuple\_id $\not\in$ hs then}\\
    \hspace*{1em}\texttt{\hspace{6em}insert getEndpoint.tuple\_id into hs;}\\
    \hspace*{1em}\texttt{\hspace{6em}break;}\\
    \hspace*{1em}\texttt{\hspace{4em}else}\\
    \hspace*{1em}\texttt{\hspace{6em}remove getEndpoint.tuple\_id from hs;}\\
    \hspace*{1em}\texttt{while not isFinished;}

\item \texttt{isFinished}:\\
    \hspace*{1em}\texttt{return iterator.isFinished;}
\end{itemize}
}
\vspace{-.5cm}

\subsection{Second Start Iterator}
\label{sec:secondstart}


\begin{itemize}
\footnotesize
\item \texttt{SecondStartIterator(iterator)}:\\
    \hspace*{1em}\texttt{this.iterator = iterator;}\\
    \hspace*{1em}\texttt{this.hs = new HashSet;}\\
    \hspace*{1em}\texttt{while getEndpoint.type = start and}\\
    \hspace*{1em}\texttt{\phantom{while }getEndPoint.tuple\_id $\not\in$ hs}\\
    \hspace*{1em}\texttt{do}\\
    \hspace*{1em}\texttt{\hspace{2em}moveToNextEndpoint;}

\item \texttt{getEndpoint}:\\
    \hspace*{1em}\texttt{return this.endpoint;}
\item \texttt{moveToNextEndpoint}:\\
    \hspace*{1em}\texttt{do}\\
    \hspace*{1em}\texttt{\hspace{2em}iterator.moveToNextEndpoint;}\\
    \hspace*{1em}\texttt{\hspace{2em}if getEndpoint.type = start then}\\
    \hspace*{1em}\texttt{\hspace{4em}if getEndpoint.tuple\_id $\in$ hs then}\\
    \hspace*{1em}\texttt{\hspace{6em}remove getEndpoint.tuple\_id from hs;}\\
    \hspace*{1em}\texttt{\hspace{6em}break;}\\
    \hspace*{1em}\texttt{\hspace{4em}else}\\
    \hspace*{1em}\texttt{\hspace{6em}insert getEndpoint.tuple\_id into hs;}\\
    \hspace*{1em}\texttt{while not isFinished;}

\item \texttt{isFinished}:\\
    \hspace*{1em}\texttt{return iterator.isFinished;}
\end{itemize}

\end{document}